\documentclass[prb,showpacs,twocolumn,superscriptaddress]{revtex4}
\usepackage{graphics}
\usepackage{graphicx}
\usepackage{dcolumn} 
\usepackage{bm}
\usepackage{float}
\usepackage{epsfig}
\begin{document}
\def\rhov{{\mbox{\boldmath{$\rho$}}}}
\def\tauv{{\mbox{\boldmath{$\tau$}}}}
\def\Deltav{{\mbox{\boldmath{$\Delta$}}}}
\def\Lambdav{{\mbox{\boldmath{$\Lambda$}}}}
\def\Thetav{{\mbox{\boldmath{$\Theta$}}}}
\def\Psiv{{\mbox{\boldmath{$\Psi$}}}}
\def\Phiv{{\mbox{\boldmath{$\Phi$}}}}
\def\sigmav{{\mbox{\boldmath{$\sigma$}}}}
\def\alphav{{\mbox{\boldmath{$\alpha$}}}}
\def\xiv{{\mbox{\boldmath{$\xi$}}}}
\def\oh{{\scriptsize 1 \over \scriptsize 2}}
\def\ot{{\scriptsize 1 \over \scriptsize 3}}
\def\of{{\scriptsize 1 \over \scriptsize 4}}
\def\tf{{\scriptsize 3 \over \scriptsize 4}}

\title{Landau Theory of Tilting of Oxygen Octahedra in Perovskites}

\author{A. B. Harris}

\affiliation{Department of Physics and Astronomy, University of
Pennsylvania, Philadelphia PA 19104}
\date{\today}
\begin{abstract}
The list of possible commensurate phases obtained from the parent
tetragonal phase of Ruddlesden-Popper systems, A$_{n+1}$B$_n$C$_{3n+1}$
for general $n$ due to a single phase transition involving the reorienting of
octahedra of C (oxygen) ions is reexamined using a Landau expansion.
This expansion allows for the nonlinearity of the octahedral rotations
and the rotation-strain coupling.
It is found that most structures allowed by symmetry are inconsistent
with the constraint of rigid octahedra which dictates the form of
the quartic terms in the Landau free energy. For A$_2$BC$_4$ our analysis
allows only 10 (see Table III) of the 41 structures listed by
Hatch {\it et al.} which are allowed by general symmetry arguments.
The symmetry of rotations for RP systems with $n>2$ is clarified.
Our list of possible structures in Table VII excludes many structures
allowed in previous studies.
\end{abstract}
\maketitle

\section{INTRODUCTION} 

\begin{figure}
\begin{center}
\includegraphics[width=8.5 cm]{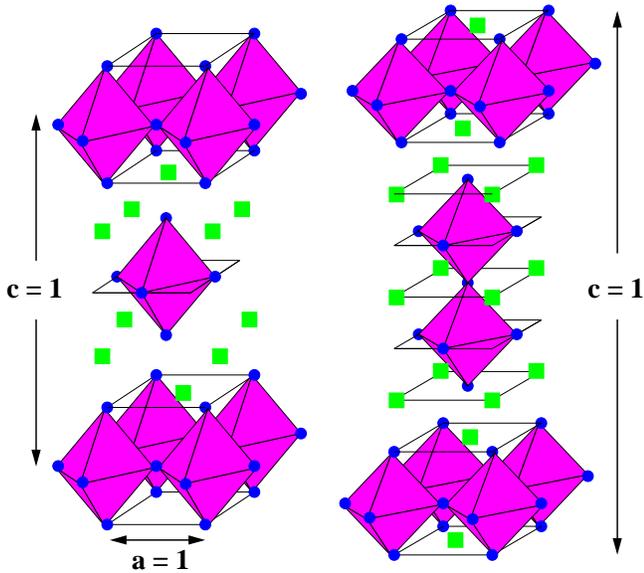}
\caption{(Color online) \label{RPFIG} The high symmetry (body centered
tetragonal) parent structure of A$_2$BO$_4$ (left) and A$_3$B$_2$O$_7$ (right).
The green squares are A ions.  The B ions are at the centers of the oxygen
(blue dots) octahedra.}
\end{center}
\end{figure}

The Ruddlesden-Popper (RP) compounds[\onlinecite{RP}] are layered
perovskites having the chemical formula
A$_{n+1}$B$_n$C$_{3n+1}$ and which consist of two slabs
of BC$_6$ octahedra per conventional unit cell.  Each slab consists of
$n$ layers of corner sharing octahedra of F's or O's.  These
systems either are or can be considered to be developed
via one or more structural transitions 
from the high symmetry tetragonal parent structure
shown in Fig. \ref{RPFIG} for the cases exemplified by
K$_2$MgF$_4$ ($n=1)$ and Ca$_3$Mn$_2$O$_7$ ($n=2$).
We will refer to the RP system with $n=1$ as RP214 and to that
with $n=2$ as RP327.

The RP systems exhibit many interesting technological
properties such as high $T_c$ superconductivity[\onlinecite{HTC}],
colossal magnetoresistance,[\onlinecite{CMR}]
metal insulator transitions,[\onlinecite{ARG}]
and coupled ferroelectric and magnetic order.[\onlinecite{LOB,CF,ABHPRB}]
Many of these properties depend sensitively on the distortions
from the ideal tetragonal I4/mmm structure (see Fig. \ref{RPFIG})
of space group \#139 [numbering of space groups is from
Ref. \onlinecite{ITC}] which appear at structural phase
transitions.[\onlinecite{STRUC1,STRUC2,STRUC3,STRUC4,STRUC5}]
Accordingly, the accurate characterization of their structure is
essential to reach a detailed understanding of their properties.
Such an understanding can potentially lead to the
fabrication of new systems with enhanced desired properties.
It is therefore not surprising that one of the most celebrated
theoretical problems in crystallography is to list the possible
structures that can result from a single structural phase transition
in which the (usually oxygen) octahedra are cooperatively reoriented
under the constraint that they are only distorted weakly (in a sense
made precise below).  One of the earliest works to address this
question was that of Glazer[\onlinecite{GLAZER}] who analyzed
possible structural distortion from the cubic parent structure
of CaTiO$_3$. It turned out that a few of the structures he
found did not actually satisfy the constraint of not distorting
the octahedra.[\onlinecite{WOODWARD}]
For the RP214 systems the two principal approaches to this problem
which have been used are a) a direct enumeration of likely 
structures[\onlinecite{ALEKS}] and b) the use symmetry.[\onlinecite{HANDS}]
This last approach utilizes a very useful computer
program[\onlinecite{COMP}] to generate the isotropy subgroups of
Ref. \onlinecite{ISOTROPY}. In this way Hatch {\it et al.}
[\onlinecite{HANDS}] gave a listing for the RP214 structure
of possible phase transitions involving distortions at various
high symmetry wave vectors. This listing was shown to be 
consistent with the revised results of method a).[\onlinecite{HANDS}] This
important work has stood unquestioned for over a decade.[\onlinecite{PHASE}]
Here we show that most of structures listed in Refs. \onlinecite{HANDS}
and \onlinecite{PHASE} for the RP214 systems are inconsistent with the 
constraint of rigid octahedra.   To implement this constraint, we
assume that the spring constants for distortion of the octahedra
are larger than the other spring constants of the lattice by a factor
of $\lambda$.  Most of our results are obtained to leading order in
$1/\lambda$, which we regard as an expansion parameter.  This constraint 
causes the quartic terms in in the Landau free energy to assume a form
which is less general than allowed by symmetry.[\onlinecite{FN6}]
In some cases this constraint causes us to reject structures which have
undistorted sublattices, a situation which is counterintuitive, since 
it is analogous to having a magnetic system simultaneously having ordered
and disordered sublattices.  Even for structures we our analysis allows,
it is inevitable that in the structural phase transition the octahedra
will undergo small (of order $1/\lambda$) distortions, which are
observed.[\onlinecite{ZHOU}]

Briefly this paper is organized as follows.  In Sec. II we enumerate the
high symmetry wave vectors of the distortions we will consider and
we discuss the role of the quartic terms in  the Landau expansion
in determining the detailed nature of the distortions.  Here we also
develop the nonlinear constraint induced by the rigidity of the octahedra.
In Sec. III we apply these ideas to enumerate the possible structures
which are allowed via a single phase transition involving a distortion 
at these high symmetry wave vectors for the RP214 structure.
In Sec. IV we extend the treatment to the analogous RP327 (A$_3$B$_2$C$_7$)
bilayer system. In Sec. V we use our results for $n=1$ and $n=2$ RP systems
to obtain results for the RP systems A$_{n+1}$B$_n$C$_{3n+1}$ consisting of
$n$-layer slabs (with $n$ finite).  In Sec. VI we discuss and summarize our results.

\section{GENERAL PRINCIPLES} 

\subsection{OVERVIEW}

We will analyze possible distortions from the parent tetragonal system
using a Landau-like formulation in which we write the free energy $F$ as
\begin{eqnarray}
F &=& \frac{1}{2} \sum_{k,l} A_{k,l}(T) X_k X_l + {\cal O} (X^4) \ ,
\end{eqnarray}
where $X_k$ is a component of an ionic displacement.  A structural phase
transition
occurs at a temperature $T_0$ when an eigenvalue of the matrix ${\bf A}$
becomes zero.  (For $T>T_0$ all the eigenvalues of ${\bf A}$ are positive.)
If the zero (critical) eigenvalue is $N$-fold degenerate, then for $T$ 
near $T_0$ one has
\begin{eqnarray}
F \sim \frac{a}{2} (T-T_0) \sum_{k=1}^N Q_k^2 \ ,
\end{eqnarray}
where $Q_k$ is the amplitude of the $k$th linear combination of $X$'s given by
the $k$th critical eigenvector of ${\bf A}$.  As we shall see, higher order
(in $Q$) corrections to the free energy in the cases of interest involve only
even powers of the $Q$'s.

\begin{figure}
\begin{center}
\includegraphics[width=5.5 cm]{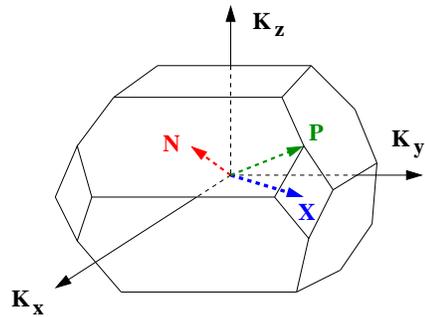}
\caption{(Color online) \label{BZ}
The first Brillouin zone for RP systems.  There are two inequivalent 
$X=(1/2,1/2,0)$ points, four inequivalent ${\bf N}$ points, and
two inequivalent ${\bf P}$ points. (Wave vectors are ``equivalent" if 
their difference involves an integer number of reciprocal lattice
vectors ${\bf G}$.)}
\end{center}
\end{figure}

As is customary ({\it e. g.} see Ref. \onlinecite{HANDS}), we will
restrict attention to the cases when $Q$ is a superposition of displacements
associated with the star of the high symmetry wave vectors
${\bf X}=(1/2,1/2,0)$, 
of ${\bf N}=(1/2,0,1/2)$, or of ${\bf P}=(1/2,1/2,1/2)$.[\onlinecite{FN1}]
These vectors are on faces of the first Brillouin zone as shown in 
Fig. \ref{BZ}.  The reciprocal lattice vectors are
\begin{eqnarray}
{\bf G}_1 &=& (-1,1,1) \ , \hspace{0.4 in}
{\bf G}_2 = (1,-1,1) \ , \nonumber \\
{\bf G}_3 &=& (1,1,-1) \ , 
\end{eqnarray}
Instead of dealing with irreducible representations (irreps),
we will develop the free energy for the most general structure which can
be constructed using the angular distortions at the wave vectors of the
star of ${\bf X}$, ${\bf N}$, or ${\bf P}$.

\begin{figure}
\begin{center}
\includegraphics[width=8.5 cm]{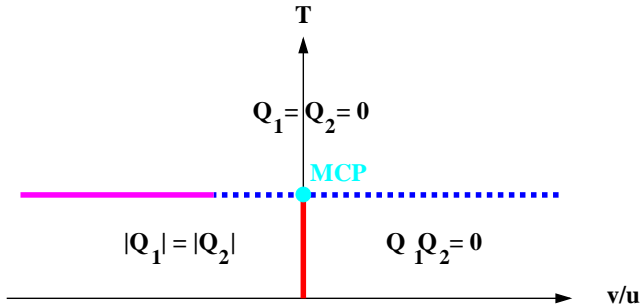}
\caption{(Color online) \label{TRIC} The mean-field phase
diagram[\onlinecite{AA}]
(schematic) for the free energy of Eq. (\ref{EQCC}), showing
the multicritical point (MCP) at $v=0$. The dashed line indicates
a continuous phase transition and the solid line a discontinuous one.}
\end{center}
\end{figure}

To see how the form of the fourth order potential affects possible structural 
distortions, consider a system with two order parameters $Q_1$ and $Q_2$
related by symmetry, for which the free energy assumes the form
\begin{eqnarray}
F &=& (T-T_0) [Q_1^2 + Q_2^2] + u [Q_1^2+Q_2^2]^2 + v Q_1^2Q_2^2 
\label{EQCC} \end{eqnarray}
up to fourth order in $Q$ with $u>0$.
As the temperature is lowered through the value $T_0$ the
nature of the ordering depends on the sign of $v$.  (See the phase diagram
of Fig. \ref{TRIC}.) If $v$ is positive, then ordering has either $Q_1$ or
$Q_2$ zero.  If $-4u<v<0$, ordering occurs with $|Q_1|=|Q_2|$.  At
the multicritical point[\onlinecite{MCP,AA}] where
$v$ is zero (and also a similar sixth order anisotropy vanishes) one can
have ordering in an arbitrary direction of order parameter ($Q_1$-$Q_2$)
space. Alternatively, analysis of terms of higher order than $Q^4$ shows
that in extreme limits ordering can occur in an arbitrary
direction in $Q_1$-$Q_2$ space.  Note that if we invoke {\it only} the
symmetry properties of the system, there is no constraint on
the parameters $u$ and $v$, in which case the analysis of Ref.
\onlinecite{HANDS} would apply.  However,
as we will explain below, the picture of the lattice
as consisting of oxygen octahedra implies a special form
of the quartic terms with $u>0$ and $v=-2u$, so that in most cases
only structures with $|Q_1|=|Q_2|$ will actually occur.

\subsection{Octahedral Constraint}

We now discuss the possible distorted configurations as constrained by the
rigidity of the oxygen (or C ion) octahedra. In describing the
distortions we use a notation similar to that of Ref. \onlinecite{HANDS}
and \onlinecite{PHASE} which we deem more convenient for the RP systems
than the widely used Glazer notation[\onlinecite{GLAZER}] for
pseudocubic perovskites.  For a single RP layer, we will first consider
the rotation of octahedra through an angle $2\theta$ about the tetragonal axis.
Giving the angle of rotation of on octahedron at the origin fixes the
rotation angles of all other octahedra in this layer.  In the Glazer
notation this would be specified as $a^0a^0b^+$ or $a^0a^0b^-$, where
the superscripts 0 indicate zero rotation about the axes $x$ and $y$,
and $b$ (implicitly equal to $2\theta$) is the rotation angle about the
third ($z$) axis.  The superscript $+$ or $-$ on $b$ tells how the
angle varies as we move from one layer perpendicular to $z$ to the
next layer.  Here, since there is only one layer, this superscript is
meaningless.  More generally, for RP systems we have two slabs to consider,
there would be two sets of Glazer symbols, one for each slab.
However, we will give a simpler symbol which applies when the star of the wave
vector is specified.  (The Glazer symbol implicitly specifies the wave vector 
by the array of superscripts.) 

\begin{figure}
\begin{center}
\includegraphics[width=7.5 cm]{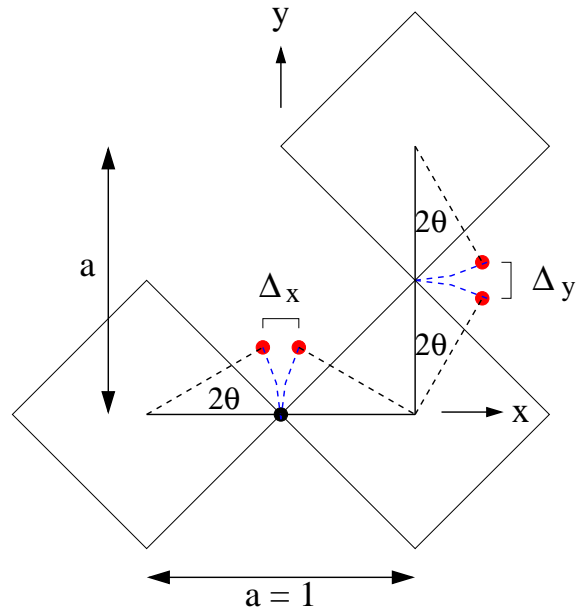}
\caption{(Color online) \label{ROTTH} The octahedral constraint for
interlocking $\theta$-rotations about the tetragonal $z$ axis. Here
$\Delta_x=\Delta_y = 2a\theta^2$, where $\Delta_\alpha$ is the mismatch
that has to be absorbed either by a distortion of the octahedron or by
a macroscopic strain when the octahedron is undistorted.}
\end{center}
\end{figure}

For a rotation about the tetragonal
$z$ axis, the situation is that shown in Fig. \ref{ROTTH}.  There
one sees that a vertex common to two adjacent octahedra would, if
the vertices were considered to be separate vertices for the two
octahedra, become two closely, but distinct, separated points.
In order to recover the common vertex, the two points would have
to coalesce, which would require octahedral distortions of $\Delta_x/2$ and
$\Delta_y/2$. However, it is possible for the lattice to relax,
so that this mode would take place {\it without} any distortion of
the octahedra.  This relaxation involves microscopic strains along
the $x$ and $y$ axes, to account for the $\Delta$ displacements.
In the presence of microscopic strains $\epsilon_{xx}$ and $\epsilon_{yy}$,
one has for RP214
\begin{eqnarray}
\Delta_{x,n} &=& a [2 \theta_n^2 + \epsilon_{xx}] \ , \nonumber \\
\Delta_{y,n} &=& a [2 \theta_n^2 + \epsilon_{yy}] \ ,
\end{eqnarray}
where $\Delta_{\alpha ,n}$ is the value of $\Delta_\alpha$ for the
$n$th slab and $\theta_n$ is the value of $\theta$ for the $n$th
of the 2 slabs in the RP214 system. Therefore for RP214 the free energy per
octahedron for the octahedrally constrained $\theta$-rotated structure
contains the term
\begin{eqnarray}
F(\theta_1, \theta_2) &=& c_\theta a^2 \lambda \sum_{k=1}^2 \left[
\left( 2 \theta_k^2 + \epsilon_{xx} \right)^2
+ \left( 2 \theta_k^2 + \epsilon_{yy} \right)^2 \right] \ ,
\label{EQ66} \end{eqnarray}
where we have introduced the expansion parameter $\lambda$ which is the ratio of the
stiffness coefficient for distorting an octahedron to other stiffness coefficients
of the lattice and $c_\theta$, and below $c_\phi$, are constants of order
unity.  Most of our results will carried only to leading order in $1/\lambda$.
Corrections higher order in $1/\lambda$ are discussed in the Appendix.  Note
that at order $1/\lambda$ it is inevitable that the octahedra will in fact
be distorted.

\begin{figure} [h!]
\begin{center}
\includegraphics[width=8.6 cm]{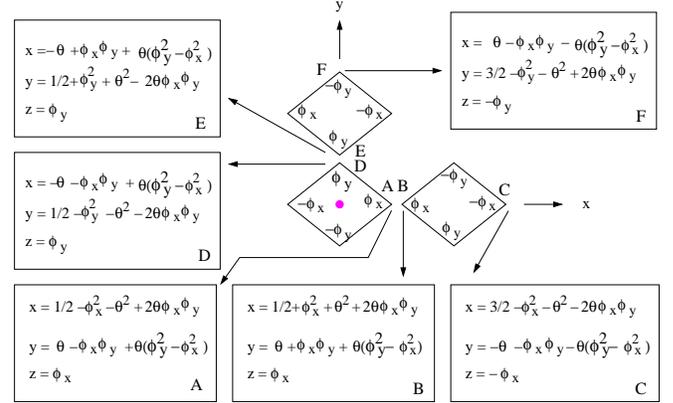}
\caption{\label{NONAB} (Color online) The displacements of the vertices
for the star of ${\bf X}$ including nonlinear contributions.  The
displacements of the lower left octahedron at the origin are
given in terms of the expansion parameters of Eq. (\ref{EQEXP}). The
$\phi$ parameters at each vertex are shown.  The $\theta$ displacements
are only given in the formulas.  Because the wave vectors of the star of
${\bf X}$ are $(1/2,1/2,0)$ and $(1/2,-1/2,0)$, the displacement
of vertex C is related to that of vertex A by changing the signs of all
the order parameters.  The displacement of vertex F is obtained similarly
from vertex D.  The displacement of vertex B is obtained from that of
vertex C by inversion about $(1,0,0)$ and that of vertex E
from that of vertex F by inversion about $(0,1,0)$.  The origin is 
indicated by the filled magneta circle.}
\end{center}
\end{figure}

We need to generalize Eq. (\ref{EQ66}) to allow for the rotation of octahedra
about the tetragonal $x$ and $y$ axes.
In the seminal work of Ref. \onlinecite{GLAZER} we are reminded that
the group of rotations is nonabelian, i. e. rotations about different
axes do not commute with one another.  Here we discuss a simple
nonlinear treatment of small rotations.  We define the orientation
of an octahedron by three variables, $\theta$, $\phi_x$, and $\phi_y$ which
correspond to small rotations about the tetragonal axes.[\onlinecite{AXES}]
These three variables correspond to the three
Euler angles needed to specify the orientation of a rigid body when
the rotation from the undistorted state is small. This situation is somewhat
similar to the spin wave expansion in which one introduces transverse
spin components to describe the rotation of a three dimensional spin.
A simple way to deal with this situation is to express the orientation
of the vectors from the center of the octahedron to the equatorial
vertices in terms of their transverse displacements as
\begin{eqnarray}
{\bf r}_1 &=& \left( \sqrt{1/4 - y_1^2 - z_1^2}, y_1, z_1 \right) \ ,
\nonumber \\
{\bf r}_2 &=& \left( x_2, \sqrt{1/4 - x_2^2 - z_2^2},  z_2 \right) \ ,
\nonumber \\
{\bf r}_3 &=& \left( - \sqrt{1/4 - y_3^2 - z_3^2}, y_3, z_3 \right) \ ,
\nonumber \\
{\bf r}_4 &=& \left( x_4, - \sqrt{1/4 - x_4^2 - z_4^2}, z_4 \right) \ ,
\end{eqnarray}
where ${\bf r}_1$, ${\bf r}_2$, ${\bf r}_3$, and ${\bf r}_4$ are
the rotated positions of the vertices whose respective original locations
were $(1/2,0,0)$, $(0,1/2,0)$, $(-1/2, 0,0)$,
and $(0,-1/2, 0)$.  Clearly, to retain the octahedral shape (with the
center of mass fixed) we require that ${\bf r}_3=-{\bf r}_1$ and
${\bf r}_4=-{\bf r}_2$.  We wish to incorporate the octahedral constraint
to leading order in  the transverse displacements. This constraint leads to
\begin{eqnarray}
|{\bf r}_1 \pm {\bf r}_2|^2 &=& \frac{1}{2} =
\left( x_2 \pm \sqrt{\frac{1}{4} - y_1^2 - z_1^2} \right)^2
\nonumber \\ && + \left( y_1 \pm \sqrt{\frac{1}{4} - y_2^2 + z_2^2} \right)^2
+ ( z_1 \pm z_2 )^2 
\end{eqnarray}
so that
\begin{eqnarray}
0 &=& x_2 \sqrt{\frac{1}{4} - y_1^2 -z_1^2} + y_1 \sqrt{\frac{1}{4} - x_2^2
-z_2^2 } + z_1 z_2 \ .
\end{eqnarray}
This gives
\begin{eqnarray}
0&=& x_2 \left[ \frac{1}{2} - y_1^2 - z_1^2 \right] + y_1 \left[
\frac{1}{2} - x_2^2 - z_2^2 \right] + z_1 z_2 + {\cal O}(q^5)\ ,
\end{eqnarray}
where $q$ is one or more of the variables. Thus
\begin{eqnarray}
\frac{x_2+y_1}{2} &=& - z_1 z_2 + x_2(y_1^2 +z_1^2) + y_1(x_2^2+z_2^2) +
{\cal O} (q^5) \ .
\end{eqnarray}
We transform from the variables $x_2$ and $y_1$ to $\theta$ and
$\delta \theta$:
\begin{eqnarray}
x_2 &=& - \theta + \delta \theta \ , \hspace{0.2 in}
y_1 = \theta + \delta \theta \ ,
\end{eqnarray}
so that
\begin{eqnarray}
\delta \theta &=& - z_1 z_2 + [- \theta + \delta \theta][(\theta+ \delta
\theta)^2 + z_1^2] \nonumber \\
&& + [ \theta + \delta \theta] [ (- \theta + \delta \theta)^2 + z_2^2] 
+ {\cal O}(q^5)\ ,
\end{eqnarray}
which gives
\begin{eqnarray}
\delta \theta &=& - z_1 z_2 + \theta (z_2^2-z_1^2) + {\cal O} (q^4)
\end{eqnarray}
To make contact with the body of the paper replace $z_1$ by $\phi_x$ and
$z_2$ by $\phi_y$, to get
\begin{eqnarray}
x_2 &=& - \theta - \phi_x \phi_y + \theta (\phi_y^2 - \phi_x^2) + {\cal O}(q^4)
\nonumber \\
y_1&=&\theta - \phi_x \phi_y + \theta (\phi_y^2 - \phi_x^2) + {\cal O}(q^4) \ 
\nonumber \\
x_1 &=& \frac{1}{2} - y_1^2 -z_1^2 = \frac{1}{2} - \phi_x^2 - \theta^2
+2 \theta \phi_x \phi_y + {\cal O}(q^4)
\nonumber \\
y_2 &=& \frac{1}{2} - x_2^2 -z_2^2 = \frac{1}{2} - \phi_y^2 - \theta^2
- 2 \theta \phi_x \phi_y + {\cal O}(q^4)
\label{EQEXP} \end{eqnarray}
One sees that in terms
of the variables $\theta$, $\phi_x$, and $\phi_y$, the actual positions of
the vertices are given by power series expansion in these variables, the 
lowest term of which identifies these variables directly with the
corresponding displacements.  (Of course, this expansion is only useful if 
the variables are small.)  We can use these variables to analyze the
symmetry of the free energy.  In calculating scattering cross sections
one must, of course, use the actual positions of the ions given by their
nonlinear expansion, such as that in Eq. (\ref{EQEXP}) or in Fig.
\ref{NONAB}, below.

In Fig. \ref{NONAB} we identify the displacements of the vertices for
the star of ${\bf X}$. From this figure we obtain the mismatch of
the vertex when considered to be two independent vertices of adjacent
octahedra to be
\begin{eqnarray}
x_B -x_A &=& 2 \phi_x^2 + 2 \theta^2 \ + \frac{\partial u_x}{\partial x}
\, \nonumber \\
y_B -y_A &=& 2 \phi_x \phi_y + \frac{\partial u_y}{\partial x}
\ , \nonumber \\
y_E - y_D &=& 2 \phi_y^2 + 2 \theta^2 + \frac{\partial u_y}{\partial y}
\ , \nonumber \\
x_E - x_D &=& 2 \phi_x \phi_y + \frac{\partial u_x}{\partial y}  \ ,
\end{eqnarray}
with corrections at order $q^4$. Here $u({\bf r})$ is the displacement field
whose derivatives give rise to the strain tensor.  
Thus, for the star of ${\bf X}$ we have
\begin{eqnarray}
F(\phi_{x,k}, \phi_{y,k},\theta_k) &=& c_\theta a^2 \lambda \sum_{k=1}^2
\left[ \left( 2 \phi_{x,k}^2 + 2 \theta_k^2 + \epsilon_{xx} \right)^2
\right.  \nonumber \\
&& \left. + \left( 2 \phi_{y,k}^2 + 2 \theta_k^2 + \epsilon_{yy} \right)^2
\right] \nonumber \\ && \ +  c_\phi a^2 \lambda \sum_{k=1}^2
\left( 2 \phi_{x,k} \phi_{y,k} + \epsilon_{xy} \right)^2  \ ,
\label{EQ17A} \end{eqnarray}
where $\epsilon_{xy} = [ \partial u_x/\partial y + \partial u_y/\partial x]/2$.
Note that it is necessary to invoke a macroscopic strain to obtain a
free energy of distortion which does not involve distorting the 
octahedra.[\onlinecite{GLAZER,WOODWARD}] Coupling the $\theta$ or $\phi$
variables to a translational phonon will not bring all the octahedra along
an axis closer together.


One might argue that
``Constraining the form of the free energy departs ... from the accepted
way of using symmetry in this theory.  Any author is certainly free to
postulate a modified free energy and derive consequences but the general 
appeal of this approach  is then limited ...."[\onlinecite{REF1}]
The reason this objection is invalid is that our treatment is
predicated on the fact that these RP systems consist of rigid
octahedra.  (This assumption of rigidity has been accepted by the
research community for several decades and is supported by recent first
principles calculations.[\onlinecite{PM,CF,INIG,TYFP}]) One can imagine
raising the temperature sufficiently or reducing the internal
force constants so as to violate our assumption that the quartic
potential due to intraoctahedral interactions dominates the
quadratic terms in the Landau expansion.  We refer to this limit as the
limit of ``octahedral melting." For the RP perovskites, this limit is
clearly irrelevant in practicality.  But in perovskites,
such as RP214, it is the geometry of the metal-oxygen bonds that leads
to the rigidity of the octahedron.  To ignore this physics and rely solely
on symmetry (as implied by Ref. \onlinecite{REF1}) is not sensible.
To summarize: if it is legitimate to consider the system as consisting
of rigid octahedra (as is the case for the RP perovskites),
then the elastic energy quartic in the ionic
displacements is dominated by coupling terms which arise from the
distortion of individual octahedra.  Note that the octahedra do not need
to be infinitely rigid for our argument to be valid.  They only need to
be rigid enough that the parameters of the quartic potential are not
very different from those for rigid octahedra.

\section{RP214 STRUCTURES}

\subsection{The star of \bf X}

\begin{figure}[h!]
\begin{center}
\includegraphics[width=7 cm]{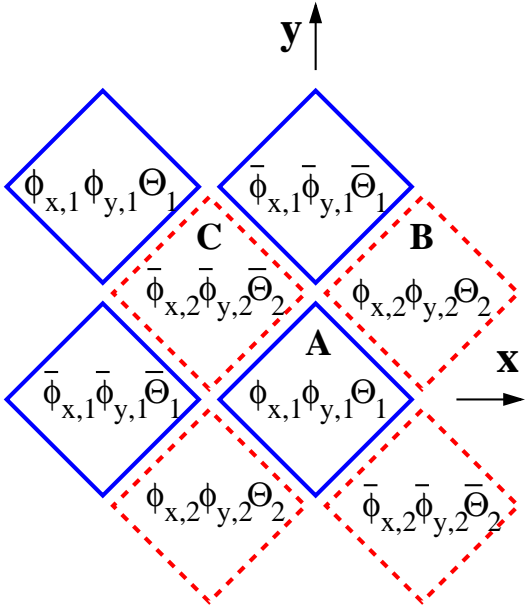}
\caption{\label{XP} (Color online) The structure of corner-sharing octahedra
for the star of ${\bf X}\equiv(1/2,1/2,0)$ and ${\bf P}\equiv (1/2,1/2,1/2)$.
The solid squares are the cross sections of octahedra in the $z=0$ plane
and the dashed squares are in the $z=1/2$ plane.
For clarity the octahedra are slightly separated instead
of sharing vertices.  Here $\phi_x$ means that the $+x$ vertex moves up by an
amount $\phi_x$ and the $-x$ vertex moves down by an amount $\phi_x$
and similarly for $\phi_y$.  Also $2\theta$ is the angle of rotation
about the $z$ axis. Here and below $\overline Q$ denotes $-Q$.
For ${\bf X}$ the structure is invariant under $z \rightarrow z+1$.  For
${\bf P}$ the variables change sign under $z \rightarrow z+1$. The
octahedral labels A, B, and C are needed for the discussion in the text.}
\end{center}
\end{figure}

As mentioned, we will develop a Landau expansion for RP214 structures
associated with the star of the wave vector, ${\bf X}$, which includes
${\bf X}_1=(1/2,1/2,0)$ and ${\bf X}_2 \equiv (1/2,-1/2,0)$ providing
that the octahedra rotate as constrained by their shared vertex.
(See Fig. \ref{XP}). In the $z=0$ plane both
${\bf X}_1$ and ${\bf X}_2$ each imply that all angular variables alternate
in sign as one moves between nearest neighbors.  This fact fixes the
values of all the $\phi$'s and $\theta$'s in the $z=0$ plane in terms of
the values for the octahedron labeled A in Fig. \ref{XP}. If we had only
the wave vector ${\bf X}_1$, then the variables for octahedron B would be
the negatives of those for octahedron A and the variables of octahedron C
would be identical to those for octahedron A.  If we had only the wave
vector ${\bf X}_2$ then the variables for octahedra B and C would be
reversed from what they were for wave vector ${\bf X}_1$.  Thus, if
we have a linear combination of the two wave vectors, the orientational
state for the plane $z=1/2$ is characterized by assigning arbitrary
values to the variables of octahedron B relative to which the values of
all the other variables in that plane are fixed. Thus Fig. \ref{XP}
gives the most general structure associated with the star of ${\bf X}$.

Therefore we write the elastic free energy for the RP214 structure of Fig.
\ref{XP} for the star of ${\bf X}$ as
\begin{eqnarray}
F(\phi_{x,k}, \phi_{y,k},\theta_k) &=& c_\theta a^2 \lambda \sum_{k=1}^2
\left[ \left( 2 \phi_{x,k}^2 + 2 \theta_k^2 + \epsilon_{xx} \right)^2 
\right. \nonumber \\
&& \left. + \left( 2 \phi_{y,k}^2 + 2 \theta_k^2 + \epsilon_{yy} \right)^2
\right] \nonumber \\ && \ +  c_\phi a^2 \lambda \sum_{k=1}^2
\left( 2 \phi_{x,k} \phi_{y,k} + \epsilon_{xy} \right)^2
\nonumber \\ && \ + F_2 + F_4 + F_\epsilon
+ F_{2,\epsilon} \ ,
\label{EQFFF} \end{eqnarray}
where the first terms come from Eq. (\ref{EQ17A}) and
last line contains terms of order $\lambda^0$.  Here
$F_2$ ($F_4$) is the free energy quadratic (quartic) in the angles
$\theta$ and $\phi$, $F_\epsilon$ is the strain free energy at quadratic
order, and $F_{2,\epsilon}$ is the rotation-strain coupling
which is quadratic in the rotation variables and linear in the strains.
Here we ignore quadratic terms involving modes which distort the octahedra,
since they will not be activated.  Also, as we shall see, terms coupling
$\phi$ and $\theta$ variables such as $\lambda \theta_k^2(\phi_{x,k}^2 +
\phi_{y,k}^2)$ do not affect the results because $\theta$ and $\phi$ are
not simultaneously critical. 

The free energy has to be invariant under all the symmetry operations of
the ``vacuum," which, in this case, is the parent tetragonal structure.
Accordingly, in Table \ref{TAB1} we give the effect of symmetry operations
on the variables appearing in Eq. (\ref{EQFFF}).  
Here and below, because of the octahedral constraint quartic terms of the
form $\theta_1^2 \theta_2^2$, $\phi_{x,1}^2 \phi_{y,2}^2 + \phi_{x,2}^2
\phi_{y,1}^2$, and $\phi_{x,1}\phi_{y,1}\phi_{x,2}\phi_{y,2}$ which
are allowed by symmetry (see Table I) do not appear at order $\lambda$.
(But, of course, they are present at order $\lambda^0$.
Only quartic terms which arise from the octahedral constraint at a single
vertex can appear at order $\lambda$.)
Using Table \ref{TAB1}, we see that
the quadratic terms which are invariant under the symmetry operations 
which leave the reference tetragonal structure invariant are
\begin{eqnarray}
F_2 &=& \alpha [ \phi_{x,1}^2 + \phi_{y,1}^2 + \phi_{x,2}^2 + \phi_{y,2}^2]
\nonumber \\ &&
+ 2 \beta [ \phi_{x,1} \phi_{y,2} + \phi_{x,2} \phi_{y,1}]
+  \gamma [ \theta_1^2 + \theta_2^2] \nonumber \\ &\equiv&
\frac{1}{2} [\alpha - \beta]\left[ \left( \phi_{x,1}-\phi_{y,2}\right)^2+
\left( \phi_{x,2} - \phi_{y,1} \right) \right] \nonumber \\ && \
\frac{1}{2} [\alpha + \beta]\left[ \left( \phi_{x,1}+\phi_{y,2}\right)^2+
\left( \phi_{x,2} + \phi_{y,1} \right) \right] 
+  \gamma [ \theta_1^2 + \theta_2^2] \ .
\end{eqnarray}
Also
\begin{eqnarray}
F_\epsilon &=& \frac{1}{2} \sum_{i,j} C_{ij} \epsilon_i \epsilon_j \ ,
\end{eqnarray}
in the Voigt notation[\onlinecite{VOIGT}]  where $1 \equiv (x,x)$,
$2 \equiv (y,y)$, etc.  Similarly, we use Table \ref{TAB1} to write
\begin{eqnarray}
F_{2 \epsilon} &=& \left[\epsilon_{xx}+\epsilon_{yy}\right]
\left[ a_1 \left( \theta_1^2 + \theta_2^2 \right) + a_2 \left(
\phi_{x,1}\phi_{y,2}+\phi_{x,2}\phi_{y,1}\right) \right. \nonumber \\ && \
+ a_3 \left. \left( \phi_{x,1}^2 + \phi_{y,1}^2 + \phi_{x,2}^2 
+ \phi_{y,2}^2 \right) \right]
\nonumber \\ && \ + \epsilon_{zz} 
\left[ a_4 \left( \theta_1^2 + \theta_2^2 \right) + a_5 \left(
\phi_{x,1}\phi_{y,2}+\phi_{x,2}\phi_{y,1}\right) \right. \nonumber \\ && \
+ a_6 \left. \left( \phi_{x,1}^2 + \phi_{y,1}^2 + \phi_{x,2}^2 
+ \phi_{y,2}^2 \right) \right]+a_7 \epsilon_{xy} \theta_1 \theta_2 \nonumber \\
&& \ + a_8 [\epsilon_{xx}-\epsilon_{yy}][ \phi_{x,1}^2 + \phi_{x,2}^2
- \phi_{y,1}^2 - \phi_{y,2} ] \nonumber \\ && \ +
a_9 [ \phi_{x,1} \phi_{y,1} + \phi_{x,2}\phi_{y,2}] \epsilon_{xy} \ .
\label{PHIEPS} \end{eqnarray}
We will deal with $F_4$ when it is needed. We will give analysis of the
above free energy which neglects fluctuations.  Our model has some
resemblance to that of Bean and Rodbell[\onlinecite{BEAN}] except that
the form of the free energy does not drive the system to a first order
phase transition, at least within mean field theory.  An interesting
problem would be to give a renormalization group analysis like that of
Bergmann and Halperin[\onlinecite{BIH}] to elucidate the effects of
orientation-strain coupling on the structural transitions.

\begin{table} [h!]
\caption{\label{TAB1} Effect of symmetry operations on the variables of the
stars of ${\bf X}\equiv(1/2,1/2,0)$ and ${\bf P}\equiv(1/2,1/2,1/2)$
(shown in Fig. \ref{XP}) and on the strain variables.  Here
${\cal R}_4$ is a four-fold rotation
about the tetragonal $z$-axis passing through the origin,
$m_d$ and $m_z$ are mirrors that take $x$ into $y$ and
$z$ into $-z$, respectively and $T$ is the translation (1/2,1/2,1/2).
These variables are odd under the translations $T_x=(1,0,0)$ and
$T_y=(0,1,0)$.  For the star of ${\bf X}$, $\xi=1$ and for the star of
${\bf P}$, $\xi=-1$.  Note that spatial inversion ${\cal I}$ is implicitly
included because ${\cal I} = {\cal R}_4^2 m_z$.[\onlinecite{REF2}]
In the last line $\Omega \equiv \epsilon_{xx}-\epsilon_{yy}$.}
\vspace{0.2 in}
\begin{tabular} {|| c | c| c| c| c||}
\hline  \hline 
& ${\cal R}_4$ & $m_d$ & $m_z$ & $T$ \\
\hline
$\phi_{x,1}$ & $\phi_{y,1}$ & $\phi_{y,1}$ & $-\phi_{x,1}$ & $\phi_{x,2}$ \\ 
$\phi_{y,1}$ & $-\phi_{x,1}$ & $\phi_{x,1}$ & $-\phi_{y,1}$ & $\phi_{y,2}$ \\ 
$\phi_{x,2}$ & $-\phi_{y,2}$ & $\phi_{y,2}$ & $-\xi \phi_{x,2}$ & $\xi
\phi_{x,1}$ \\ 
$\phi_{y,2}$ & $\phi_{x,2}$ & $\phi_{x,2}$ & $-\xi \phi_{y,2}$ &
$\xi \phi_{y,1}$\\ 
\hline
$\theta_1$ & $\theta_1$ & $-\theta_1$ & $\theta_1$ & $\theta_2 $ \\
$\theta_2$ & $-\theta_2$ & $-\theta_2$ & $\xi \theta_2$ & $\xi \theta_1 $ \\
\hline
$\epsilon_{xy}$ & $-\epsilon_{xy}$ & $\epsilon_{xy}$ & $\epsilon_{xy}$ & 
$\epsilon_{xy}$ \\
$\Omega$ & $-\Omega$ & $-\Omega$ & $\Omega$ & $\Omega$ \\
\hline \hline \end{tabular}
\end{table}

The structural phase transitions which we are investigating arise
when {\it one} of the channels becomes unstable, i. e. when $\gamma$ or
$\alpha - |\beta|$ passes through zero. (As in Ref. 
\onlinecite{HANDS}, we reject multicritical points where more
than one channel simultaneously becomes unstable.)  

\subsubsection{$\theta$ distortion}

\begin{figure}[h!]
\begin{center}
\includegraphics[width=7 cm]{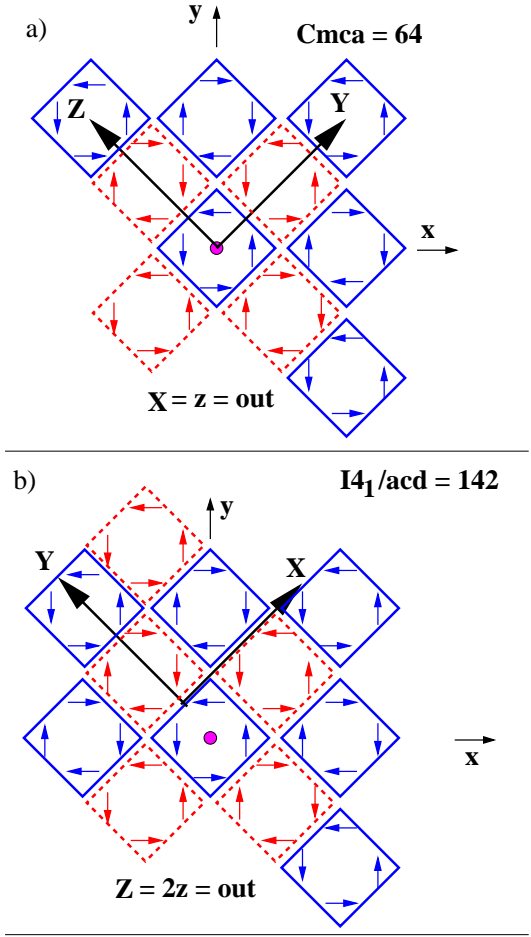}
\caption{\label{THROT} (Color online) As Fig. \ref{XP}.
The structure of corner-sharing octahedra
obtained by $\theta$ rotations either for ${\bf X}\equiv(1/2,1/2,0)$ (top)
and ${\bf P}\equiv (1/2,1/2,1/2)$ (bottom). The arrows indicate the
displacement of the oxygens in the equatorial plane.  Here and in the figures
below the original tetragonal axes are labeled by lower case letters and
those of the distorted structure are labeled by capital letters.  The
magenta dot represents the tetragonal origin. In the top (bottom) panel
$X=(0,0,1)_t$ ($Z=(0,0,2)_t$), where the subscript indicates  that components
are taken in the original tetragonal system.  In the top (bottom) panel the
distortion is unchanged (changes sign) for $z \rightarrow z+1$.
In the top (bottom) panel the new origin is at $z=0$ ($z=3/4$).}
\end{center}
\end{figure}

For instance, when only $\gamma$ becomes negative, then 
\begin{eqnarray}
\phi_{x,1}&=& \phi_{x,2}=  \phi_{y,1}= \phi_{y,2}=0 \ ,
\end{eqnarray}
so that
\begin{eqnarray}
F(\theta)  &=& c_\theta 
a^2 \lambda \sum_{k=1}^2 \left[ \left( 2 \theta_k^2 +
\epsilon_{xx} \right)^2 + \left( 2 \theta_k^2 + \epsilon_{yy}
\right)^2 \right] \nonumber \\ && \
- \frac{1}{2} |\gamma | [ \theta_1^2 + \theta_2^2] + F_4 (\theta) 
+ F_\epsilon + F_{2, \epsilon} \ ,
\end{eqnarray}
where $F_4(\theta)$ are quartic terms in $\theta$ of order $\lambda^0$.
To leading order in $1/\lambda$, when $F(\theta)$ is minimized, one finds that
\begin{eqnarray}
\epsilon_{xx} &=& \epsilon_{yy} = -2 \theta_1^2 = -2 \theta_2^2 \ .
\label{EQ7} \end{eqnarray}
(Corrections to this result at leading order in $1/\lambda$ are
studied in Appendix A. In particular, there are no such corrections to
the relation $\theta_1^2 = \theta_2^2$.) As we discuss in a moment,
all choices of signs for the $\theta$'s in Eq. (\ref{EQ7}) lead to
equivalent structures, one of which is shown in Fig.  \ref{THROT}a.

Now we identify the space group for the above $\theta$ distortion.  The
${\bf X}$- structure has generators[\onlinecite{FN4}] $(X\pm 1/2,Y+1/2,Z)$,
$(X,Y,Z+1)$,
$(\overline X, \overline Y, \overline Z)$, $(X, \overline Y, \overline Z)$,
and $(\overline X + 1/2, \overline Y , Z + 1/2)$.  In determining the space
group from these generators,[\onlinecite{FN8}] it is useful to realize that
the structures we find must form a subset of those listed in Ref.
\onlinecite{HANDS}.  We thus identify the space group of the structure
of Fig. \ref{THROT}a as D$_{2h}^{18}$ or Cmca (\#64).
Cmca (64), one of the three $\theta$-dependent structures of irrep $X_2^+$
for the star of ${\bf X}$ which are listed in Ref. \onlinecite{HANDS}.
However, we do not allow the other two structures of Ref. \onlinecite{HANDS},
the first of which (D$_{2h}^9$ or Pbam=\#55) has, according to Table III of
Ref. \onlinecite{HANDS}, $|\theta_1| \not= |\theta_2|$,
with $\theta_1\theta_2 \not= 0$ and the second of which (D$_{4h}^5$ or
P4/mbm=\#127) has, according to Table III of Ref. \onlinecite{HANDS},
one sublattice distorted and the other not distorted, so
that $\theta_1 \theta_2=0$ (see Fig. \ref{TRIC}). The problem with
these structures is that to avoid distorting the octahedra we had
to invoke a uniform strain which only relieves the distortions in
the two slabs when the distortions in the two slabs are the same.
Thus when $\theta_1^2 \not= \theta_2^2$, there is an unavoidable
distortion energy of the octahedra of order $\lambda$.

Note also that the term in $F_{2\epsilon}$ proportional to
$\epsilon_{xy} \theta_1 \theta_2$ in combination with $F_\epsilon$ leads to
\begin{eqnarray}
\epsilon_{xy} &=& - a_7 \theta_1 \theta_2 / c_{44} \ ,
\end{eqnarray}
This distortion is consistent with orthorhombic symmetry and
with the orientation of the orthorhombic axes shown in Fig.
\ref{THROT}. Rotating the crystal by 90$^{\rm o}$ 
(${\cal R}_4$) changes the sign of $\theta_1\theta_2$ and thus
changes the sign of $\epsilon_{xy}$, as one would expect.

In comparison to other ordering transitions we can make an analogy between
the order parameters which govern the distortion from the parent tetragonal
phase and the order parameters in, say, a magnetic system. In this
formulation the distortion of the parent lattice in perovskites is
analogous to the development of long-range magnetic order. Having
a distortion only within one sublattice of the RP system is thus
analogous to having magnetic order only on some sublattices.  Although one
can have ordered systems which have some disordered components, they
differ from the present case.  For instance, the orientational phase II of 
solid methane (CD$_4$) consists of a unit cell having six orientationally
ordered methane
molecules and two completely disordered molecules.[\onlinecite{PRESS}]
In that case, the site symmetry of the disordered molecules is high
enough that the effective field from the ordered molecules vanishes.
Furthermore, the interaction energy $E$ between two disordered molecules
is much less than $kT$, even at the lowest temperature at which phase II
exists, so that they do not cooperatively order.  Superficially, the situation
here is similar to that for solid methane.  For instance, suppose one builds
up the distorted structure plane by plane for wave vectors in the star of
${\bf X}$ or ${\bf P}$.  The first plane would have
$\theta=\theta_0$, say.  Moving to the second plane we note a frustration
due to the four-fold rotation, ${\cal R}_4$ which implies that the energy
is invariant against changing the sign of $\theta_2$. (Table \ref{TAB1}
indicates that ${\cal R}_4$ leaves $\theta_1$ fixed but takes $\theta_2$
into $-\theta_2$.)  If one considers the
third plane, there is no analogous frustration.  One will have 
$\theta_3 = \pm \theta_1$, the sign depending on the weak interaction 
between octahedra in plane \#1 and those in plane \#3. More generally,
$\theta_{n+2}= \sigma \theta_n$, where $\sigma$ can be either $+1$
or $-1$. As in quasi-two
dimensional systems, as the temperature at which the distortion becomes
unstable is approached, two dimensional correlations will become significant
and then even a weak coupling in the third dimension will lead to 
three dimensional long-range order at a single phase transition.
This situation is reminiscent of the decoupling of magnetic sublattices
in the bcc antiferromagnet.[\onlinecite{SHENDER,TANER}] The point is
that if the distortion order parameter becomes nonzero in, say, the
even numbered planes, the mechanism that led to this order would
also apply to the odd numbered planes, which would then also distort
at the same time.
The distorting of both sublattices of octahedra could only be avoided
if simultaneous distortions were strongly disfavored by the form of the
quartic potential ({\it i. e.} if $v$ of Eq. (\ref{EQCC}) were positive.)
This possibility seems unlikely and indeed our analytic treatment of
the octahedral constraint indicates that this scenario does not occur
for large $\lambda$.

\subsubsection{$\phi$ distortions}

Now drop the $\theta$ variables, so that[\onlinecite{AXE}] the
$\phi$-dependent free energy for the star of ${\bf X}$ is
\begin{eqnarray}
F(\phi) 
&=& c_\theta a^2 \lambda \sum_{k=1}^2 \left[ \left( 2 \phi_{x,k}^2
+ \epsilon_{xx} \right)^2 + \left( 2 \phi_{y,k}^2
+ \epsilon_{yy} \right)^2 \right] \nonumber \\ && \
[(\alpha - \beta )/2] \left[ ( \phi_{x,1}-\phi_{y,2})^2
+ ( \phi_{x,2}-\phi_{y,1})^2 \right]
\nonumber \\ && + [(\alpha + \beta)/2] \left[
( \phi_{x,1}+\phi_{y,2})^2
+ ( \phi_{x,2}+\phi_{y,1})^2 \right]  \nonumber \\ &&
+ c_\phi a^2 \lambda \sum_{k=1}^2 \left( 2 \phi_{x,k} \phi_{y,k}
+ \epsilon_{xy} \right)^2 + F_4 (\phi) \ ,
\end{eqnarray}
where $F_4(\phi)$ are the terms quartic in $\phi$ which are proportional
to $\lambda^0$. When $F(\phi)$ is minimized for large $\lambda$, we get
\begin{eqnarray}
\epsilon_{xx}= -2\phi_{x,1}^2 = -2\phi_{x,2}^2 \hspace {0.2 in} \Rightarrow
\hspace{0.2 in} \phi_{x,2} = \pm \phi_{x,1} \ ,
\label{EPS1} \end{eqnarray}
\begin{eqnarray}
\epsilon_{yy}= -2\phi_{y,1}^2 = -2\phi_{y,2}^2 \hspace{0.2 in} \Rightarrow
\hspace{0.2 in} \phi_{y,2} = \pm \phi_{y,1} \ ,
\label{EPS2} \end{eqnarray}
and
\begin{eqnarray}
\epsilon_{xy} = - 2 \phi_{x,1} \phi_{y,1} = - 2 \phi_{x,2} \phi_{y,2} \ .
\label{EPS3} \end{eqnarray}

\begin{figure} [h!]
\includegraphics[width=6.6 cm]{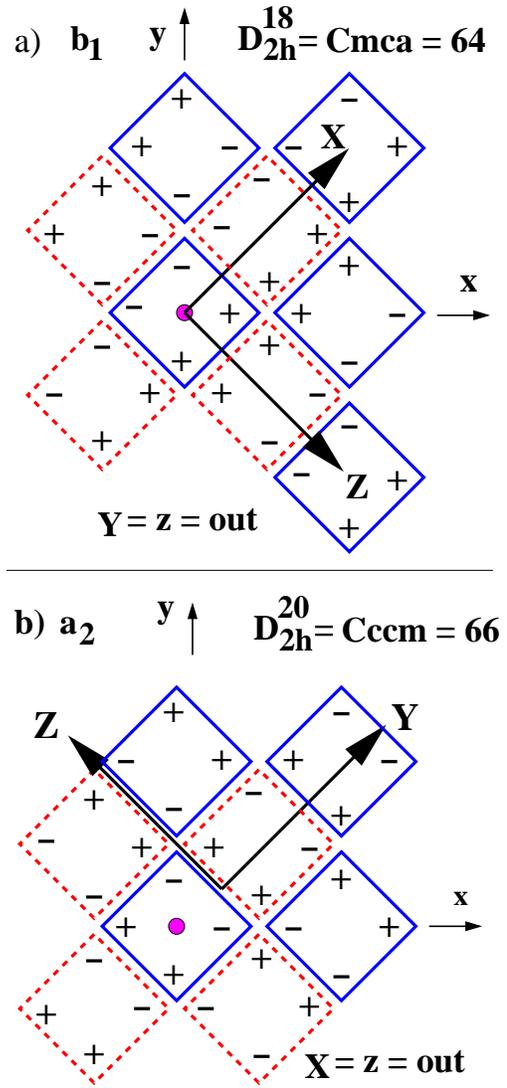}
\vspace{0.1 in}
\caption{\label{XFIG} (Color online)
As Fig. \ref{XP} for the star of ${\bf X}$ (with invariance under
$z \rightarrow z+1$).  $x$, $y$, and $z$ are the tetragonal axes and
and $X$, $Y$, and $Z$ are the conventional lattice vectors after distortion.
The filled magenta circle is the tetragonal origin. 
$\Psi \equiv [\phi_{x,1}, \phi_{y,1},\phi_{x,2},\phi_{y,2}]=$
$[1 \overline 1 1 \overline 1]$ for a) and 
$[\overline 1 \overline 1 \overline 1 \overline 1 ]$ for b).[\onlinecite{FN35}]
The irreps are given in Table \ref{YES}.  The new origin has $z=0$ for a)
and $z=1/4$ for b).  The third new axis vector is $[001]_t$ in terms of the
original tetragonal coordinates.}
\end{figure}

There are four possible directions of the ordering vector
$\Psi \equiv [\phi_{x,1}, \phi_{y,1} , \phi_{x,2} , \phi_{y,2} ]$
depending on whether or not $\alpha-\beta$ becomes critical
(zero) before $\alpha+\beta$ and the choice of signs
in Eqs. (\ref{EPS1}) and (\ref{EPS2}).
When $\alpha-\beta$ is critical (so that $\phi_{x,1}+\phi_{y,2}=
\phi_{x,2}+\phi_{y,1}=0$), then $\Psi$ is proportional to either
$a_1 =  [11 \overline 1 \overline 1]$  or
$b_1 =  [1 \overline 1 1 \overline 1 ]$.
If $\alpha+\beta$ is critical (so that $\phi_{x,1}-\phi_{y,2}=
\phi_{x,2}-\phi_{y,1}=0$), then $\Psi$ is proportional to either
$a_2 =  [1111 ]$ or $b_2 =  [1 \overline 1 \overline 1 1]$.
In each case, the two choices are equivalent:[\onlinecite{FN34}]
${\cal R}_4 b_n = a_n$. Figure  \ref{XFIG}
shows a representative of these solutions for each case.[\onlinecite{FN35}]

Finally, we identify the space groups of the structures of Fig.  \ref{XFIG}.
The generators of $b_1$ are
$(X\pm 1/2, Y+1/2,Z)$,$(X,Y,Z+1)$,
$(\overline X, \overline Y, \overline Z)$, $(X,\overline Y,\overline Z)$,
and $(\overline X +1/2, \overline Y, 1/2+Z)$ and those of $a_2$ are
$(X \pm 1/2, Y+1/2,Z)$, $(X,Y,Z+1)$, $(\overline X , \overline Y ,\overline Z)$,
$(\overline X,\overline Y, Z)$, and $(X, \overline Y, 1/2+ \overline Z)$.
From these generators we identify the space groups as indicated in Fig.
\ref{XFIG}.  From Eq. (\ref{PHIEPS}) we see from the term in $a_8$ that 
$\epsilon_{xx}=\epsilon_{yy}$ for both these structures.
The term in $a_9$ indicates that $\epsilon_{xy} \not= 0$,
consistent with the orientation of the coordinate axes shown in
Fig. \ref{XFIG}. 

Note that, in comparison to Ref. \onlinecite{HANDS}, our formulation
does not allow the structures of space groups Pccn (D$_{2h}^{10}=$ \#56)
and Pnnn (D$_{2h}^2=$ \#48).  From Table III of Ref. \onlinecite{HANDS}
one sees that Pccn has $\phi_{x,1}=\phi_{y,2}=a$ and
$\phi_{y,1}=\phi_{x,2}=b$, with $a \not= b$ and Pnnn has
$\phi_{x,1}=-\phi_{y,2}=a$ and $\phi_{y,1}=-\phi_{x,2}=b$, with
$a \not= b$.  As before, to relieve the distortion of the octahedra
via strains implies that the order parameters have the same magnitude
on both sublattices.
Similarly we do not allow space groups P4$_2$/ncm (D$_{4h}^{16}=$
\# 138) and P4$_2$/nnm (D$_{4h}^{12}=$ \# 134) which have (in one setting)
$\phi_{x,1}=\phi_{y,2}=0$ and $|\phi_{x,2}|=|\phi_{y,1}|$.
Our analysis of the space groups listed in Ref.
\onlinecite{HANDS} is summarized in Tables \ref{NO} and \ref{YES}. 

\begin{table}[h!]
\caption{\label{NO}
Space groups of Ref. \onlinecite{HANDS}
for the stars if ${\bf X}$, ${\bf N}$, and ${\bf P}$ which we do not allow
for RP214's.  For the column headings ``S`` stands for the Schoenflies symbol,
``H-M" stands for the short Hermann-Maugin symbol as given in Ref.
\onlinecite{ITC}, and \# is the number of the space group in 
Ref. \onlinecite{ITC}. For otherwise identical space groups, the
footnotes give the basis vectors of the unit cell in terms of the original
tetragonal coordinates. The irrep labels are from Ref. \onlinecite{HANDS}.}

\vspace{0.2 in}
\begin{tabular} {||c|c c c|| c c c ||}
\hline
Irrep & S & \#& H-M& S & \# & H-M \\
\hline
X$_2^+$ & D$_{4h}^5$ & 127 & P$_4$/mbm & D$_{2h}^9$ & 55 & Pbam \\
X$_3^+$ & D$_{2h}^{10}$ & 56 & Pccn &
D$_{4h}^{16}$ &138 &  P4$_2$/ncm \\
X$_4^+$ & D$_{2h}^2$ & 48 & Pnnn &
D$_{4h}^{12}$ & 134 & P4$_2$/nnm \\
N$_1^+$ & C$_{2h}^1$ & 10 & P2/m & C$_i^1$ & 2 & P${\overline 1}^{***}$ \\
N$_1^+$ & D$_{2h}^{28}$ & 74 & Imma$^*$ & D$_{2h}^{25}$ & 71 & Immm \\
N$_1^+$ & C$_{2h}^6$ & 15 & C2/c & C$_{2h}^3$ & 12 & C2/m$^{\dagger}$ \\
N$_1^+$ & C$_{2h}^3$ & 12 & C2/m$^{\dagger \dagger}$
& D$_{4h}^{19}$ & 141 & I4$_1$/amd\\
N$_1^+$ & D$_{2h}^{19}$ & 65& Cmmm &
D$_{4h}^{17}$ & 139 & I4/mmm \\
P$_4$ &D$_{4h}^{18}$ & 140 & I4/mmm & D$_{2d}^{10}$ & 122 & I$\overline 4$2d \\
P$_5$ &  D$_{2h}^{28}$ & 74 & Imma$^{**}$ & D$_{2h}^{24}$ & 69 & Fmmm \\
P$_5$ & S$_4^2$ & 82 & I$\overline 4$ & C$_{2v}^{22}$ & 46 & Ima2  \\
P$_5$ & C$_{2v}^{19}$ & 43 & Fdd2 & C$_{2v}^{18}$ & 42 & Fmm2  \\
P$_5$ & D$_2^9$ & 24 & I2$_1$2$_1$2$_1$ & D$_2^8$ & 23 & I222  \\
P$_5$ & D$_2^7$& 22 & F222 & C$_{2h}^6$ & 15 & C2/c \\
P$_5$ & C$_{2h}^3$ & 12 & C2/m$^{\dagger \dagger \dagger}$ & C$_2^3$ & 5 & C2 \\
\hline \hline
\end{tabular}

\vspace{0.2 in} \noindent
$^*$: $(200),\ \ (020), \ \ (002)$.  \hspace{0.2 in}
$^{**}$: $(002),\ \ (1 \overline 1 0), \ \ (110)$.

\noindent
$^{***}$: \hspace{0.2 in} $(\overline 1 11), \ \ (\frac{1}{2}
\overline{\frac{1}{2}} \frac{1}{2}, \ \ (11 \overline 1)$.
\hspace{0.2 in} $\dagger$: $(00\overline 2),\ \ 
(\overline 2 \overline 20),\ \ (\overline 111)$.

\noindent
$\dagger \dagger$:  $(\overline 20\overline 2),\ \
(0\overline 20),\ \ (0,0,2)$.
\hspace{0.2 in} \noindent
$\dagger \dagger \dagger$: $(020),\ \ (002), \ \ (1 \overline 1 0)$.
\noindent
\end{table}

\begin{table} [h!]
\caption{\label{YES}
As Table \ref{NO}. Space groups of Ref. \onlinecite{HANDS} for the stars
of ${\bf X}$, ${\bf N}$, and ${\bf P}$ which we do allow for RP214's.
Under ``Var's" we give the variables active in the mode and under
``Fig" we give the number of the illustrative figure.}

\vspace{0.2 in}
\begin{tabular} {||cc |c c c |c|| }
\hline
Irrep &Var's& S & \# & H-M & Fig \\
\hline
X$_2^+$ & $\theta$ & D$_{2h}^{18}$ & 64 & Cmca &7a \\
X$_3^+$ & $\phi_x$,$\phi_y$ & D$_{2h}^{18}$ &64&Cmca&8a \\
X$_4^+$ & $\phi_x$,$\phi_y$ & D$_{2h}^{20}$ & 66 & Cccm& 8b \\
N$_1^+$ & $\phi_y$ & C$_{2h}^3$ & 12 & C2/m$^\dagger$&10b \\
N$_1^+$ & $\phi_x$,$\phi_y$ & C$_{2h}^3$ & 12 & C2/m$^{\dagger \dagger}$ &10c \\
N$_1^+$ & $\phi_x$, $\phi_y$ & C$_i^1$ & 2 & P$\overline 1^{**}$ & 10a \\
P$_4$ & $\theta$ & D$_{4h}^{20}$ & 142 & I4$_1$/acd & 7b \\ \hline \hline
P$_5^*$ & $\phi_x$,$\phi_y$ &D$_{2h}^{26}$ & 72 & Ibam & 11a \\
P$_5^*$ & $\phi_x$,$\phi_y$ & C$_{2h}^{6}$ & 15
& C2/c & 11c \\
P$_5^*$ & $\phi_y$ & D$_{2h}^{24}$ & 70 & Fddd & 11b \\
\hline \hline
\end{tabular}

\vspace{0.2 in} $\dagger$: $(0\overline 11),\ \ (100),\ \ (011)$.

\noindent
$\dagger \dagger$:  $(002),\ \ (220),\ \ (\overline{\frac{1}{2}}\frac{1}{2}
\overline{\frac{1}{2}})$.

\noindent
* The wave vector may be close to, but not exactly at, the star of ${\bf P}$.

\noindent
$**$:  $(\overline 111), \ \ (1 \overline 1 1),\ \ (11 \overline 1)$.

\end{table}

\subsection{The star of \bf N}

\begin{figure}[h!]
\begin{center}
\includegraphics[width=7 cm]{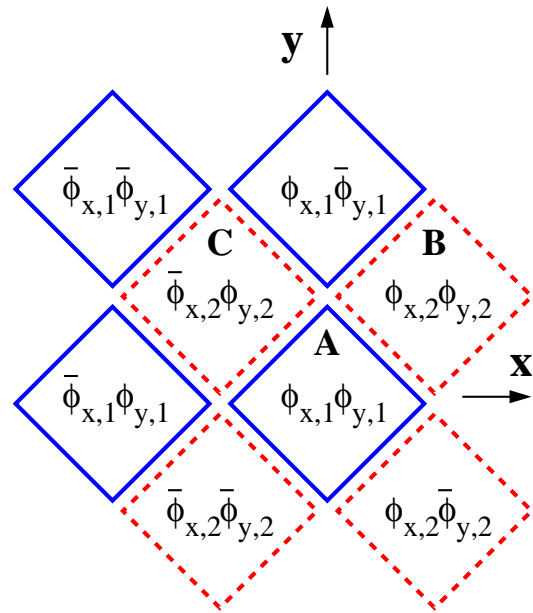}
\vspace{0.1 in}
\caption{\label{N} (Color online) As Fig. \ref{XP}.  The structure
of corner-sharing octahedra for the star of ${\bf N}$.  The variables
change sign under $z \rightarrow z+1$.}
\vspace{0.1 in}
\end{center}
\end{figure}

\begin{table} [h!]
\caption{\label{NTAB} As Table \ref{TAB1}, but for the RP327 variables of
the star of ${\bf N}\equiv (1/2,0,1/2)$ (shown in Fig. \ref{N}).
$T\equiv (1/2,1/2,1/2)$, $T_x \equiv (1,0,0)$, and $T_y \equiv (0,1,0)$.
All variables are odd under $T_z \equiv (0,0,1)$.}
\vspace{0.2 in}
\begin{tabular} {|| c | c| c| c| c|c|c||}
\hline  \hline 
& ${\cal R}_4$ & $m_d$ & $m_z$ & $T$ & $T_x$ & $T_y $\\
\hline
$\phi_{x,1}$ & $\phi_{y,1}$ & $\phi_{y,1}$ & $-\phi_{x,1}$ & $\phi_{x,2}$ &
$-\phi_{x,1}$ & $\phi_{x,1}$ \\ 
$\phi_{y,1}$ & $-\phi_{x,1}$ & $\phi_{x,1}$ & $-\phi_{y,1}$ & $\phi_{y,2}$ &
$\phi_{y,1}$ & $- \phi_{y,1}$ \\ 
$\phi_{x,2}$ & $\phi_{y,2}$ & $\phi_{y,2}$ & $\phi_{x,2}$ & $\phi_{x,1}$ &
$-\phi_{x,2}$ & $\phi_{x,2}$ \\ 
$\phi_{y,2}$ & $\phi_{x,2}$ & $\phi_{x,2}$ & $\phi_{y,2}$ & $\phi_{y,1}$ &
$\phi_{y,2}$ & $-\phi_{y,2}$ \\ 
\hline \hline \end{tabular}
\end{table}

Similarly, we construct the most general structure for the star of
${\bf N}\equiv(1/2,0,1/2)$ whose wave vectors are
${\bf N}_1 \equiv (1/2,0,1/2)$, ${\bf N}_2=(-1/2,0,1/2)$,
${\bf N}_3=(0,1/2,1/2)$, and ${\bf N}_4=(0,-1/2,1/2)$.  To do this
we use Fig. \ref{N}. Consider first the situation in the $z=0$ plane.
Note that $\phi_x$ has to alternate in sign as we move along $x$.
This means that $\phi_x$ is associated with a linear combination of
${\bf N}_1$ and ${\bf N}_2$ distortions.  Since $N_{1,y}=N_{2,y}=0$, we
see that $\phi_x$
must be independent of $y$.  Similar reasoning indicates that
$\phi_y$ alternates along $y$ and is therefore associated with a
linear combination of ${\bf N}_3$ and ${\bf N}_4$ distortions.  Therefore
$\phi_y$ is independent of $x$. These wave vectors
do not support nonzero values of $\theta$.  We have thereby fixed all
the values of the variables in the $z=0$ plane in terms of those of
octahedron A.  Now consider the situation in the $z=1/2$ plane.  Suppose
we have a linear combination of ${\bf N}_1$ and ${\bf N}_2$ which gives
rise to $\phi_{x,1}$ for octahedron A. If we had only ${\bf N}_1$, then
the variables for octahedron B would be $-\phi_{x,1}$ and $-\phi_{y,1}$,
whereas if we had ${\bf N}_2$, then these variables would be $\phi_{x,1}$ and
$\phi_{y,1}$.  As for the case of the star of ${\bf X}$, we conclude that
all the variables in the second layer are fixed in terms of the arbitrary
variables of octahedron B, so that Fig. \ref{N} characterizes the most
general structure arising from the star of ${\bf N}$.  The effect of
symmetry operations on these variables is given in Table \ref{NTAB}.

We need to establish the analog of Fig. \ref{NONAB} for the star of
${\bf N}$.  First of all, here we can set $\theta=0$.  Also, note that
for sites B and C, the sign of $\phi_y$ is reversed for the star of ${\bf N}$
in comparison to that for the star of ${\bf X}$.  For site A, ${\bf r}_A$
for the star of ${\bf N}$ is as in Fig. \ref{NONAB}. However, now
\begin{eqnarray}
{\bf r}_B &=& (1/2+\phi_x^2, - \phi_x \phi_y, \phi_x) \ .
\end{eqnarray}
Therefore $y_B-y_A=z_B-z_A=0$ and $x_B-x_A=2 \phi_x^2$.
Likewise, for site D Fig. \ref{NONAB} applies equally for the
star of ${\bf N}$.  However, for sites E and F the sign of $\phi_x$
is reversed for the star of ${\bf N}$ from what it was for the star of
${\bf X}$.  So 
\begin{eqnarray}
{\bf r}_E &=& (-\phi_x \phi_y, 1/2+\phi_y^2 , \phi_y) \ ,
\end{eqnarray}
so that $x_E-x_D=z_E-z_D=0$ and $y_E-y_D=2 \phi_y^2$.

We now write the Landau expansion for the star of ${\bf N}$.
Taking account of the symmetries of Table \ref{NTAB} and the
preceding discussion, we obtain
the relevant terms in the free energy for the star of ${\bf N}$ to be
\begin{eqnarray}
F_N &=& c_\theta a^2 \lambda \left[ 
\left( 2 \phi_{x,1}^2+\epsilon_{xx} \right)^2
+\left( 2 \phi_{y,1}^2+\epsilon_{yy} \right)^2 \right. \nonumber \\
&& \ + \left. \left( 2 \phi_{x,2}^2+\epsilon_{xx} \right)^2
+ \left( 2 \phi_{y,2}^2+\epsilon_{yy} \right)^2\right]
\nonumber \\ && \
+ \frac{\alpha}{2} \left[ \phi_{x,1}^2  + \phi_{y,1}^2 + \phi_{x,2}^2 
+ \phi_{y,2}^2 \right] + F_{2\epsilon} + F_4 \ ,
\label{NEQ} \end{eqnarray}
where
\begin{eqnarray}
F_4 &=& \frac{u}{4} \left[ \left( \phi_{x,1}^2 + \phi_{y,1}^2 \right)^2
+ \left( \phi_{x,2}^2 + \phi_{y,2}^2 \right)^2 \right] \nonumber \\
&& + v \left( \phi_{x,1}^2 \phi_{y,1}^2 + \phi_{x,2}^2 \phi_{y,2}^2
\right)
+ w \left( \phi_{x,1}^2 \phi_{x,2}^2 + \phi_{y,1}^2 \phi_{y,2}^2 \right)
\nonumber \\ && \
+ x \left( \phi_{x,1}^2 \phi_{y,2}^2 + \phi_{y,1}^2 \phi_{x,2}^2 \right)
\ .
\label{F4EQ} \end{eqnarray}
To leading order in $1/\lambda$ the minima of $F_N$ occur for
\begin{eqnarray}
\epsilon_{xx} = - 2\phi_{x,1}^2 = -2\phi_{x,2}^2 \hspace{0.2 in} \Rightarrow
\hspace{0.2 in} \phi_{x,2}=\pm \phi_{x,1} \nonumber \\
\epsilon_{yy} = -2\phi_{y,1}^2 = -2\phi_{y,2}^2 \hspace{0.2 in} \Rightarrow
\hspace{0.2 in} \phi_{y,2}=\pm \phi_{y,1} \ .
\end{eqnarray}

The quartic terms distinguish between these solutions.  If
$\phi_x^2\equiv \phi_{x,1}^2=\phi_{x,2}^2$ and similarly for $\phi_y$,
then we have
\begin{eqnarray}
F_4 &=& 2(v-w+x) \phi_x^2 \phi_y^2 + [w+(u/2)] (\phi_x^2 + \phi_y^2)^2 \ .
\end{eqnarray}
This indicates that we can have a structural phase transition into
three classes of states.  We can have a continuous phase transition
into states of class A with $\phi_x^2=\phi_y^2$ if $(v-w+x)$ is negative
or into a state of class B with $\phi_x\phi_y=0$ if $(v-w+x)$ is positive.
In addition, we can have a phase transition to a state of class C in which
$\phi_x$ and $\phi_y$ do not assume special values if the higher order terms
cause the transition to be discontinuous.
The next step is to determine which of these solutions are inequivalent,
i. e. which are not related by a symmetry operation.[\onlinecite{FN34}]

We first show that all solutions of class A are equivalent to
one another.  Using the results given in Table \ref{NTAB} we have that
\begin{eqnarray}
\left( 1 + {\cal R}_4 + {\cal R}_4^2 + {\cal R}_4^3 \right)
[1111] &=& [ \{ \mu \nu \}11] \ ,
\end{eqnarray}
where $\{\mu \nu \}$ indicates the set of $\mu$, $\nu$ values, i. e.
$\{ \mu \nu \} = 11 + 1 \overline 1 + \overline 1 1 + \overline 1 \overline 1$.
Then
\begin{eqnarray}
T\left( 1 + {\cal R}_4 + {\cal R}_4^2 + {\cal R}_4^3 \right) [1111] &=& 
[ \overline 1 \overline 1 \{ \mu \nu \}] \equiv \Phi \ ,
\end{eqnarray}
and finally
\begin{eqnarray}
\left( 1 + {\cal R}_4 + {\cal R}_4^2 + {\cal R}_4^3 \right) \Phi
&=& [ \{\rho \tau \} \{\mu \nu\}] \ ,
\end{eqnarray}
where (apart from an arbitrary amplitude) the right-hand side of this
equation includes all vectors of class A. In class A, $\phi_x^2=\phi_y^2$,
so that $\epsilon_{xx}=\epsilon_{yy}$  and we take $[\overline 1 1 \overline 1
\overline 1]$ as its representative.  The term in Eq. (\ref{PHIEPS})
proportional to $a_9$ indicates that $\epsilon_{xy} \not= 0$ for this
structure.

Next we consider solutions of class B. From Table \ref{NTAB} note that
\begin{eqnarray}
[1 + {\cal R}_4^2][1+ T_x] [1010] &=& \sum_{\mu \nu =\pm 1}
[\mu 0 \nu 0] \equiv \Phi\ . 
\end{eqnarray}
and
\begin{eqnarray}
{\cal R}_4 \Phi &=& \sum_{\mu \nu = \pm 1} \left( [\mu 0 \nu 0 ]
+ [0 \mu 0 \nu ] \right) \ .
\end{eqnarray}
The right-hand side includes all vectors of class B.  We may take
$[010\overline 1]$ as the representative of class B.

Finally we consider solutions of class C, which are of the form
$[x , y , \pm x , \pm y ]$, where $|x| \not= |y|$ and both
are nonzero.  Using Table \ref{NTAB} we write 
\begin{eqnarray}
m_d {\cal R}_4  [x,y,x,y]&=&[x \overline y x y]
\end{eqnarray}
so that
\begin{eqnarray}
[1 + m_z] [1 + m_d {\cal R}_4] [xyxy] = \sum_{\sigma_1 = \pm 1} 
\sum_{\sigma_2 = \pm 1} (\sigma_1 x , \sigma_2 y , x, y] \ .
\end{eqnarray}
From this we conclude that all vectors of class C are equivalent
to one another, and we take their representative to be
$[\overline xy\overline x\overline y]$.

\begin{figure} [h!]
\includegraphics[width=8. cm]{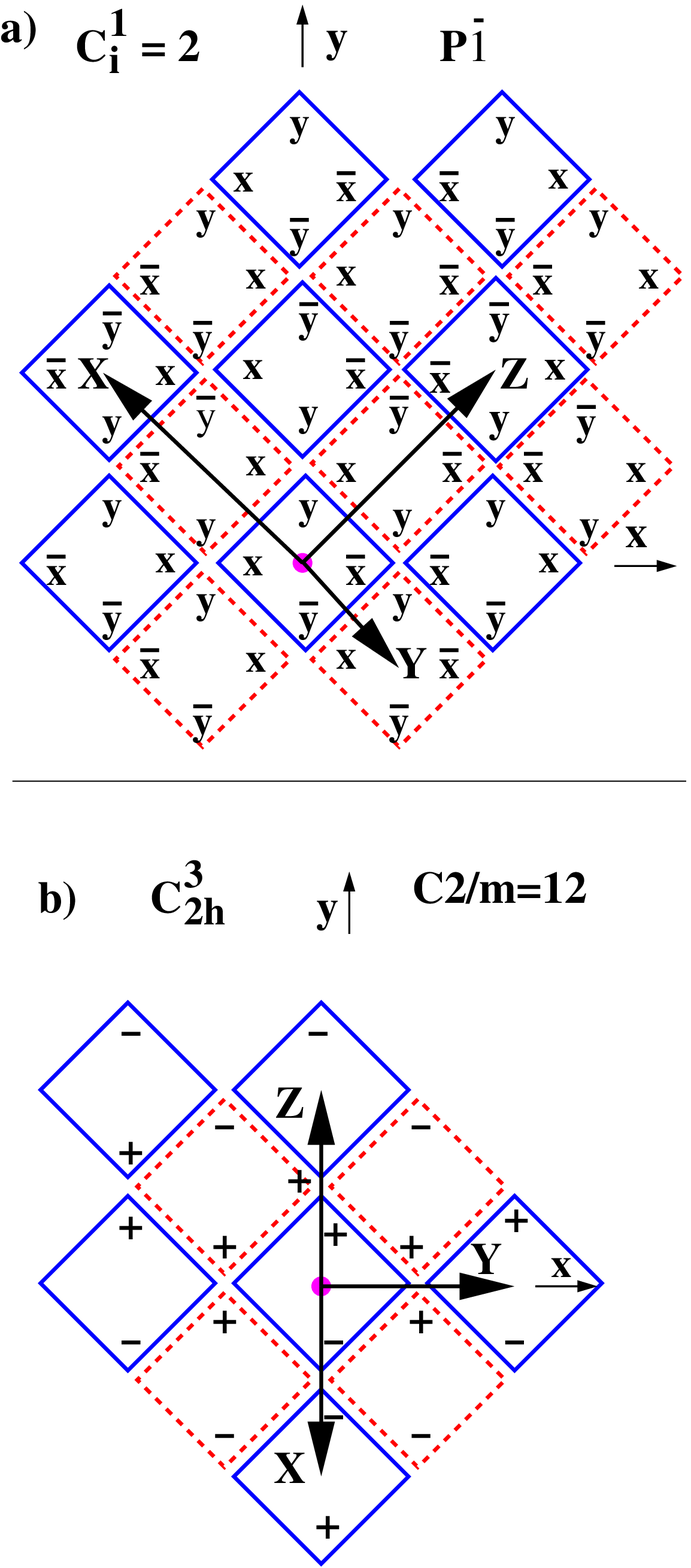}
\vspace{0.1 in}
\caption{\label{NFIG} (Color online)
As Fig. \ref{XFIG} for the star of ${\bf N}$ (with sign change under
$z \rightarrow z+1$) for a) (class C)
$[\overline x y \overline x \overline y ]$ and
b) (class B) $[01 0 \overline 1 ]$.
$x$, $y$ and $z$ are the tetragonal axes, and $X$, $Y$, and $Z$ are
$[( \overline 111),(1/2,-1/2,1/2),(11 \overline 1)]$ in a) and
$[(0\overline 1 1),(100),(011)]$ in b).  All the new origins are at 
$z=0$.}
\end{figure}

\begin{figure} [h!]
\includegraphics[width=8. cm]{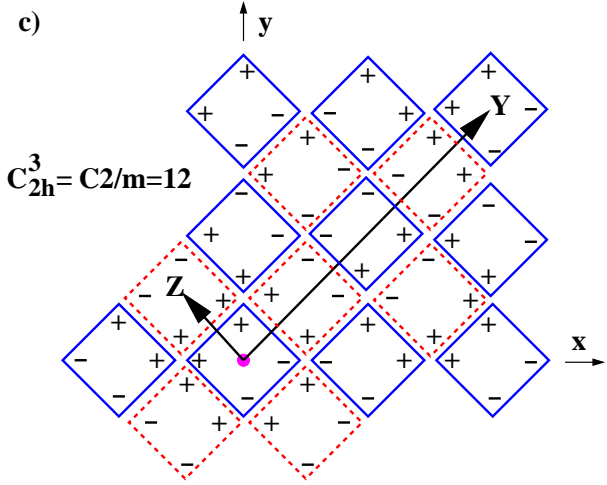}
\vspace{0.1 in}
\caption{\label{NFIGb} (Color online)
Continuation of Fig. \ref{NFIG}.  for c) (class A)
$[\overline 1 1 \overline 1 \overline 1]$.
$X$, $Y$, and $Z$ are [(002),(220)($-1/2,1/2,-1/2$)].
The new origins is at $z=0$.}
\end{figure}

Thus, in all,  we have the three allowed space groups from the star
of ${\bf N}$ shown in Figs. \ref{NFIG} and \ref{NFIGb}:
$[\overline x y \overline x \overline y]$, $[01 0 \overline 1 ]$, and
$[\overline 1 1 \overline 1 \overline 1]$.
As before, to determine the space groups of the structures we use
Figs. \ref{NFIG} and \ref{NFIGb} to write down their generators.  For
$[\overline x y \overline x \overline y ]$ the generators are
$(X+1,Y,Z)$, $(X,Y+1,Z)$, $(X,Y,Z+1)$, and $(\overline X, \overline Y,
\overline Z)$, which is C$_i^1$ (P$\overline 1 =$ \# 2).  For both
$[010\overline 1 ]$ and $[\overline 1 1\overline 1 \overline 1]$, the
generators are $(X,Y,Z+1)$, $(X+1/2,Y+1/2,Z)$, $(X+1/2, Y-1/2,Z)$,
$(\overline X , \overline Y , \overline Z )$, and
$(\overline X , Y , \overline Z )$ which is C$_{2h}^3$ (C2/m $=$ \#12).
Although these two structures belong to the same space group, they are
different because their unit cells are different (see Figs. \ref{NFIG}
and \ref{NFIGb}).
 
In Table \ref{YES} (\ref{NO}) we list the structures which are (not) allowed. 
We only allow one of the eight structures for the irrep ${\bf N}_1^+$
listed in Ref.  \onlinecite{HANDS} which occur via a discontinuous transition.
In addition, Ref. \onlinecite{HANDS} lists two space groups Cmmm (65) for
which $\Psi=[1000]$  and I4/mmm(139) for which $\Psi=[1100]$.  Both these
are inconsistent with the fourth order terms arising from the rigid 
octahedral constraint.  [I. e. they did not arise from Eq. (\ref{NEQ})].
They are also counterintuitive in that they both describe states in which
nonzero order parameters appear only on alternate planes. [Look back at
the discussion below Eq. (\ref{EQ7})].

\subsection{The star of \bf P}

The star of ${\bf P}$ consists of the vectors ${\bf P}_1 \equiv (1/2,1/2,1/2)$
and ${\bf P}_2 \equiv (1/2,-1/2,1/2)$.  The possible structures are
identical to those for the star of ${\bf X}$, except that the variables
change sign under $z \rightarrow z+1$ as indicated in Fig. \ref{XP}.
The transformation properties of these variables are given in Table
\ref{TAB1}.   The quartic terms are the same as for the star of ${\bf X}$.
However, the quadratic terms differ because of the $\xi$ factors that
appear in Table \ref{TAB1}.  The octahedral constraint assumes the same form
as for the star of ${\bf X}$ because the stars of ${\bf X}$ and ${\bf P}$
only differ in how the layers are stacked. Therefore for the star of
${\bf P}$ we have
\begin{eqnarray}
F(\phi_{x,k}, \phi_{y,k},\theta_k) &=& c_\theta a^2 \lambda \sum_{k=1}^2
\left[ \left( 2 \phi_{x,k}^2 + 2 \theta_k^2 + \epsilon_{xx} \right)^2 
\right. \nonumber \\
&& \left. + \left( 2 \phi_{y,k}^2 + 2 \theta_k^2 + \epsilon_{yy} \right)^2
\right] \nonumber \\ && \ +  c_\phi a^2 \lambda \sum_{k=1}^2
\left( 2 \phi_{x,k} \phi_{y,k} + \epsilon_{xy} \right)^2
\nonumber \\ && \ + F_2 + F_4 + F_\epsilon
+ F_{2,\epsilon} \ .
\label{EQPPP} \end{eqnarray}

\subsubsection{$\theta$ distortions}

We analyze these as before.  In the channel where $\gamma$ passes through
zero, we have
\begin{eqnarray}
F(\theta)  &=& c_\theta 
a^2 \lambda \sum_{k=1}^2 \left[ \left( 2 \theta_k^2 +
\epsilon_{xx} \right)^2 + \left( 2 \theta_k^2 + \epsilon_{yy}
\right)^2 \right] \nonumber \\ && \
- \frac{1}{2} |\gamma | [ \theta_1^2 + \theta_2^2] + F_4 (\theta) 
+ F_\epsilon + F_{2, \epsilon} \ ,
\label{PPEQ} \end{eqnarray}
so that $|\theta_1|=|\theta_2|$.  
This structure has the same degeneracy associated with the relative phase of
even and odd layers that we saw for the previous $\theta$ structures
(at the star of ${\bf X}$).  Because
$P_z=1/2$, the only $\theta$-dependent structure has $\theta_{n+2}=-\theta_n$.
The other two structures listed in
Ref. \onlinecite{HANDS} which have $|\theta_1| \not= |\theta_2|$
are not admissible due to the form of the term proportional to
$\lambda$ in Eq. (\ref{PPEQ}).  The
allowed ${\bf P}$-structure, shown in Fig. \ref{THROT}b has generators
$(X-1/2,Y+1/2,Z+1/2)$, $(X+1/2,Y-1/2,Z+1/2)$, $(X+1/2,Y+1/2,Z-1/2)$,
$(\overline X , \overline Y, \overline Z)$, 
$(X, \overline Y, \overline Z +1/2)$,
and $(\overline Y + 1/4, X+ 3/4, Z+ 1/4)$, which is space group 
D$_{4h}^{20}$ or I4$_1$/acd (\# 142).

\subsubsection{$\phi$ distortions}

Now consider the $\phi$-dependent solutions which are associated with irrep
P$_5$ according to Ref. \onlinecite{HANDS}.  Note that all the subgroups from
this irrep listed there and in Ref. \onlinecite{ISOTROPY} do not satisfy 
the Lifshitz condition.  What this
means is that the quadratic instability occurs at a wave vector that is not
fixed by symmetry[\onlinecite{YOYT}] (so that it can not be assumed to be at
the star of ${\bf P}$).  Accordingly, the wave vector can only be at
the star of ${\bf P}$ if the transition is discontinuous.

\begin{figure} [h!]
\includegraphics[width=6.0 cm]{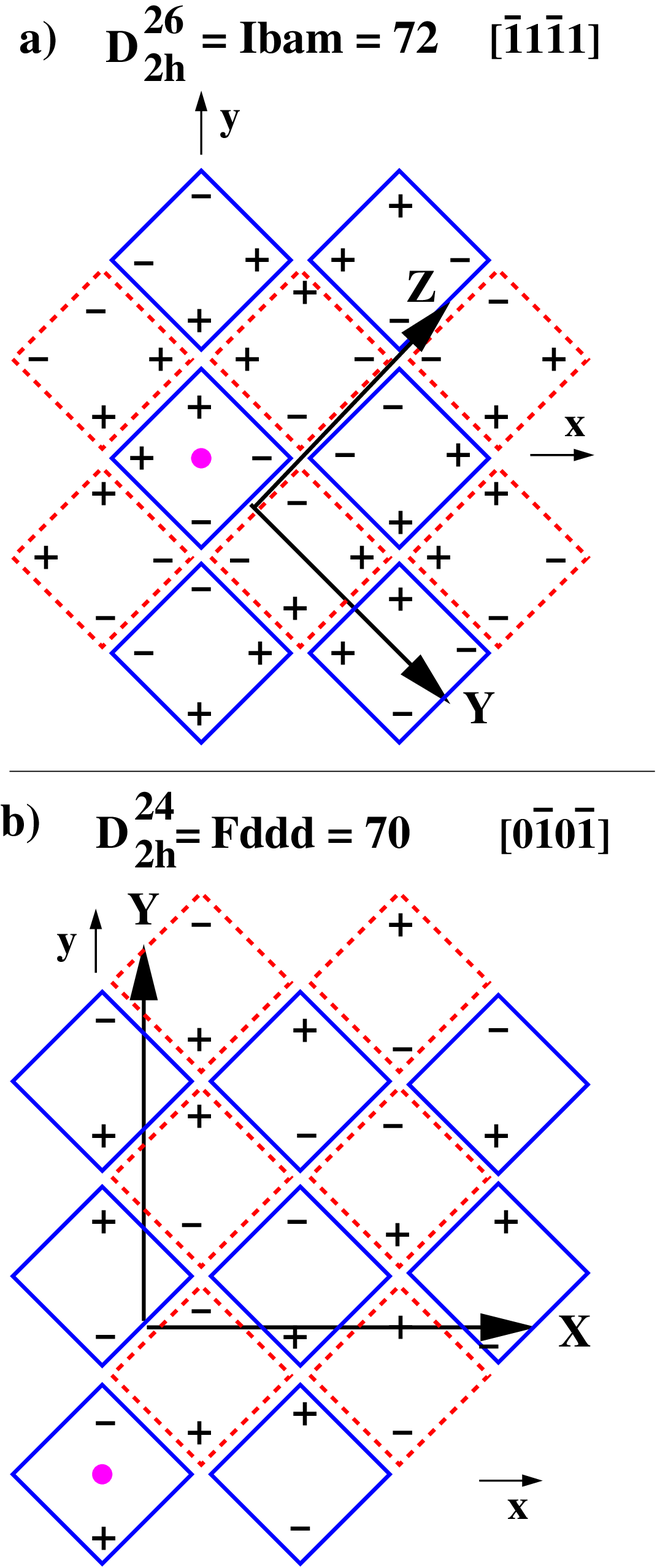}
\vspace{0.1 in}
\caption{\label{PFIG} (Color online)
As Fig. \ref{XFIG} for the star of ${\bf P}$ (variables
change sign under $z \rightarrow z+1$).  $x$, $y$, and $z$ are the tetragonal
axes and and $X$, $Y$, and $Z$ are axes of the distorted structure.
The new origins are in the $z=3/4$ plane. 
The new out-of-plane lattice vector has magnitude 
2 and is along $z$. The actual structures may involve
an incommensurate wave vector near the star of ${\bf P}$.}
\end{figure}

However, we can determine which structures with wave vectors either at
or near the star of ${\bf P}$ can be condensed.
To do this, we simply ignore the Lifshitz criterion
in our analysis of the free energy.  For this purpose we consider the
$\phi$-dependent free energy which is
\begin{eqnarray}
F &=& c_\theta a^2 \lambda \sum_{k=1}^2 \left[ \left( 2 \phi_{x,k}^2 +
\epsilon_{xx} \right)^2 \right. \nonumber \\ && \ +
\left. \left( 2 \phi_{y,k}^2 + \epsilon_{yy} \right)^2 \right]
\nonumber \\ && \ + c_\phi a^2 \lambda \sum_{k=1}^2 \left( 2 \phi_{x,k}
\phi_{y,k} + \epsilon_{xy} \right)^2 \nonumber \\ && \
+ \frac{1}{2} \alpha \left[ \phi_{x,1}^2 + \phi_{y,1}^2 + \phi_{x,2}^2
+ \phi_{y,2}^2 \right] 
\nonumber \\ && \ + F_4 + F_\epsilon + F_{2,\epsilon} \ .
\end{eqnarray}
and we consider the phase transition which occurs when $\alpha$ passes
through zero.  For large $\lambda$ we must have
\begin{eqnarray}
\epsilon_{xx} &=& - 2 \phi_{x,1}^2 = - 2 \phi_{x,2}^2
\nonumber \\ 
\epsilon_{yy} &=& - 2 \phi_{y,1}^2 = - 2 \phi_{y,2}^2
\nonumber \\ 
\epsilon_{xy} &=& -2 \phi_{x,1}\phi_{y,1} = -2 \phi_{x,2}\phi_{y,2} \ .
\end{eqnarray}
Thus we have three classes of solutions: in class A we have
$[1,0,\pm 1,0]$ and $[0,1,0,\pm 1]$.  In class B we have
$[11 \sigma \sigma ]$ and $[1 \overline 1 \sigma \overline \sigma ]$
with $\sigma=1$, class C is like  B, except that $0 < \sigma < 1$.
These three classes are associated with the different minima of
$F_4$ which assumes the same form as in Eq. (\ref{F4EQ}).
The analysis to show that all members of same class are actually
equivalent follows the previous arguments, so we will not repeat it here.
The three solutions are shown in Fig. \ref{PFIG}. We take the representatives
to be $[0 \overline 1 0 \overline 1 ]$ for class A,
$[\overline 1 1 \overline 1 1]$ for class B, and $[xyxy]$ for class C.

\begin{figure} [h!]
\includegraphics[width=6.0 cm]{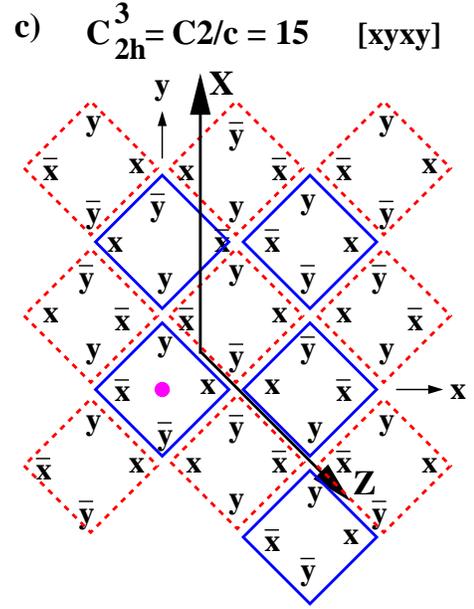}
\vspace{0.1 in}
\caption{\label{PFIGb} (Color online)
Continuation of Fig. \ref{PFIG}.  The new origin
is in the $z=1/4$ plane.  The new out-of-plane lattice vector has magnitude 
2 and is along $z$.}
\end{figure}

As before, we identify the space groups of these solutions by determining
the generators of the representatives shown in Fig. \ref{PFIG}. For
a) ($[\overline 1 1 \overline 1 1 ]$) the generators are[\onlinecite{IBAM}]
$(X, Y, \overline Z)$, $(X, \overline Y , \overline Z+1/2)$, $(\overline X, \overline Y,
\overline Z)$, $(X-1/2,Y+1/2,Z+1/2)$, $(X+1/2,Y-1/2,Z+1/2)$, 
$(X+1/2, Y+1/2, Z-1/2)$ and thus the space group is Ibam (D$_{2h}^{26}=$\#72).
For b) ($[0 \overline 1 0\overline 1])$
the generators are $(\overline X + 1/4, \overline Y + 1/4, Z$),
$(X,\overline Y +1/4, \overline Z +1/4)$, $(\overline X, \overline Y,
\overline Z)$, $(X, Y+1/2,Z+1/2)$, $(X+1/2,Y,Z+1/2)$,
$(X+1/2,Y+1/2,Z)$, and thus the space group is Fddd (D$_{2h}^{24}=$\#70).
For c) ($[xyxy])$ the generators are
$(\overline X, Y,\overline Z + \frac{1}{2} )$,
$(\overline X, \overline Y, \overline Z)$, ($X+1/2,Y+1/2,Z)$, $(X-1/2,Y+1/2,
Z)$, $(X,Y,Z+1)$ and thus the space group is C2/c (C$_{2h}^3=$\#15). 
When the Lifshitz instability is resolved, these space groups give rise
either to commensurate structures at the star of ${\bf P}$ via
a first order transitions or to structures having
incommensurate wave vectors near the star of ${\bf P}$.

\section{RP327 STRUCTURES}

\begin{figure} [h!]
\begin{center}
\includegraphics[width=7.5 cm]{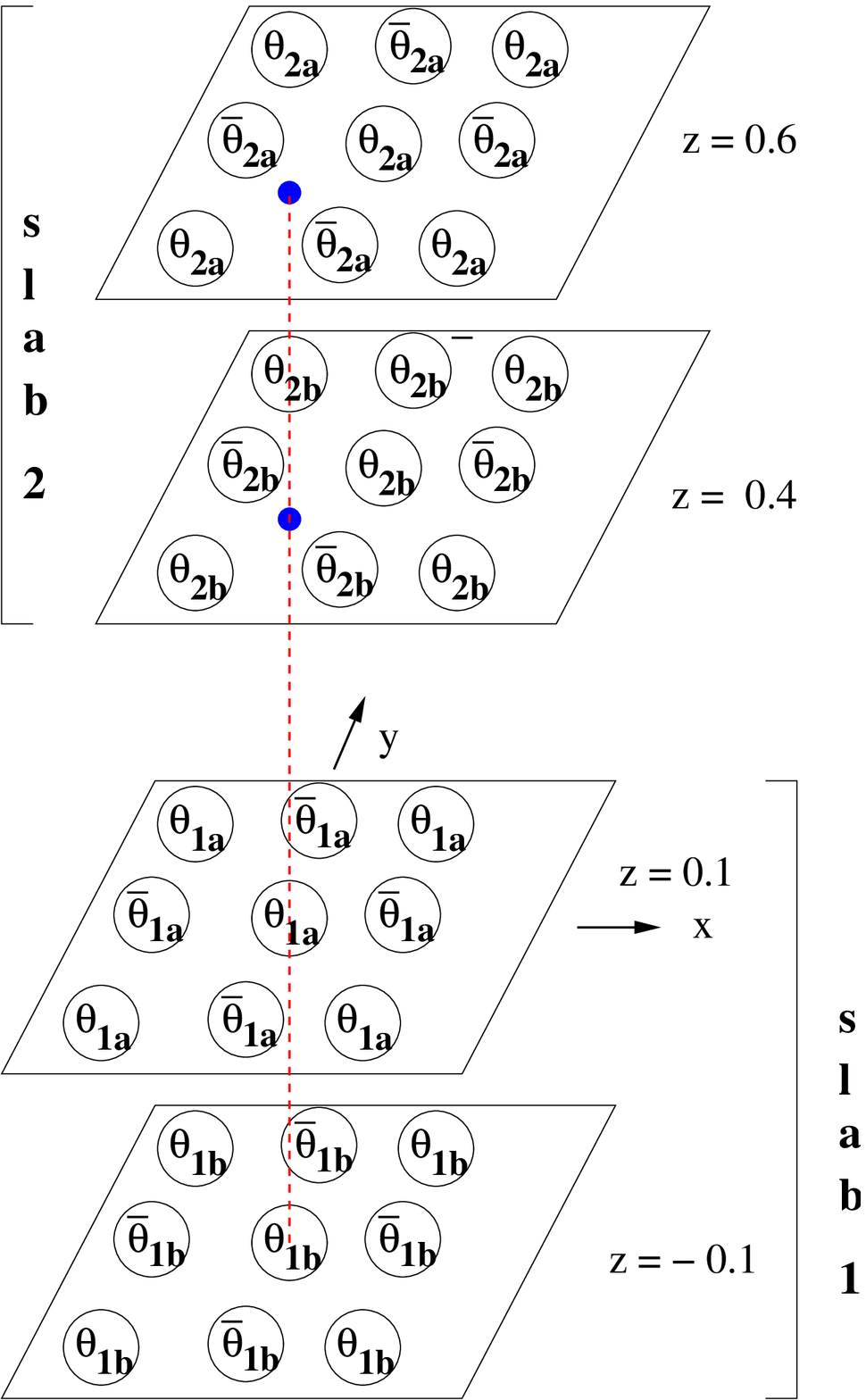}
\caption{(Color online) \label{THETA}
Values of the rotation ($\theta$) variables in different $z$-planes
for the most general such
distortion of RP327's for the stars of ${\bf X}$ and ${\bf P}$.  The (red)
dashed line is the axis about which the fourfold rotation ${\cal R}_4$ is
taken and the (blue) dots show the points $x=y=0$ in the $z=0.4$ 
and $z=0.6$ planes.  Under ${\cal R}_4$ an octahedron in slab \#1 is
taken either into itself or into an equivalent octahedron.
Under ${\cal R}_4$ an octahedron in slab \#2 is taken
into another whose rotation angle is of opposite
sign.  For the star of ${\bf X}$, $\theta_{n+2,x}=\theta_{n,x}$ and for
the star of ${\bf P}$, $\theta_{n+2,x}=-\theta_{n,x}$.}
\end{center}
\end{figure}

In this section we perform the same analysis for the $n=2$ RP
systems A$_3$B$_2$C$_{7}$ which we call the RP327 systems.
As shown in Fig. \ref{RPFIG}, these systems consist of two slabs.
Each slab consists of two layers which we label $a$ and $b$
(or 1 and 2).
If we fix the $\phi$ variables in layer $a$, the octahedral constraint
fixes each $\phi$ variable in layer $b$ to be the negative of its
nearest neighbor in layer $a$.  As a result each structure of the
RP327 system is characterized by the same number of $\phi$ variables
as its analog for the RP214 system.  In contrast, since there is no
such relation between the $\theta$ variables of layers $a$ and $b$,
we introduce variables $\theta_{n,x}$ to
describe the rotation within the $x$th layer ($x=a,b$) of the
$n$th slab ($n=1,2$), as shown in Fig. \ref{THETA}. The
transformation properties of the variables are summarized in
Table \ref{TAB327}.

\begin{table} [h!]
\caption{\label{TAB327} As Table \ref{TAB1} for the stars of ${\bf X}$ and 
${\bf P}$, except that this table is for the variables of RP327 systems. Note
that $z=0$ (about which $m_z$ is taken) is midway between layers $a$ and $b$.
$T$ is the translation $(1/2,1/2,1/2)$.}
\vspace{0.2 in}
\begin{tabular} {|| c | c| c| c| c||}
\hline  \hline 
& ${\cal R}_4$ & $m_d$ & $m_z$ & $T$ \\
\hline
$\theta_{1,a}$ & $\theta_{1,a}$ & $-\theta_{1,a}$ 
& $\theta_{1,b}$ & $\theta_{2,a}$ \\ 
$\theta_{1,b}$ & $\theta_{1,b}$ & $-\theta_{1,b}$
& $\theta_{1,a}$ & $\theta_{2,b} $ \\
$\theta_{2,a}$ & $- \theta_{2,a}$ & $-\theta_{2,a}$
& $\xi \theta_{2,b}$ & $\xi \theta_{1,a}$\\ 
$\theta_{2,b}$ & $-\theta_{2,b}$ & $-\theta_{2,b}$ &
$\xi \theta_{2,a}$ & $\xi \theta_{1,b} $ \\
\hline
$\phi_{x,1}$ & $\phi_{y,1}$ & $\phi_{y,1}$ & $\phi_{x,1}$ &
$\phi_{x,2}$ \\
$\phi_{y,1}$ & $-\phi_{x,1}$ & $\phi_{x,1}$ & $\phi_{y,1}$ &
$\phi_{y,2}$ \\
$\phi_{x,2}$ & $-\phi_{y,2}$ & $\phi_{y,2}$ & $\xi \phi_{x,2}$ &
$\xi \phi_{x,1}$ \\
$\phi_{y,2}$ & $\phi_{x,2}$ & $\phi_{x,2}$ & $\xi \phi_{y,2}$ &
$\xi \phi_{y,1}$ \\
\hline \hline \end{tabular}
\end{table}

\subsection{The star of {\bf X}}

\subsubsection{$\theta$ structures}

Using the symmetry operations of Table \ref{TAB327} we find that
the free energy of the $\theta$ structures for the star of ${\bf X}$
assumes the form
\begin{eqnarray}
F(\theta)&=& c_\theta a^2 \lambda \sum_{k=1}^2 \sum_{\alpha=a}^b
\left[ \left( 2\theta_{k\alpha}^2 + \epsilon_{xx} \right)^2 +
\left( 2\theta_{k\alpha}^2 + \epsilon_{yy} \right)^2 \right] \nonumber \\ && \
+ \frac{1}{2} a \sum_{k=1}^2 \left( \theta_{ka}^2 + \theta_{kb}^2
\right) \nonumber \\ && \
+ b \left[ \theta_{1a} \theta_{1b} + \theta_{2a} \theta_{2b} \right]
+ F_4 + F_{2 \epsilon} \ .
\end{eqnarray}
For large $\lambda$, the minima of this free energy as $(a-|b|)$ passes
through zero occur for
\begin{eqnarray}
\epsilon_{xx} &=& \epsilon_{yy} = -2 \theta_{1a}^2 = -2 \theta_{1b}^2
\nonumber \\ &=& -2 \theta_{2a}^2 = -2 \theta_{2b}^2 \nonumber \\ 
\theta_{1a} &=& - \frac{b}{|b|} \theta_{1b} \ , \hspace{0.3 in}
\theta_{2a} = - \frac{b}{|b|} \theta_{2b} \ .
\end{eqnarray}

\begin{figure} [h!]
\begin{center}
\includegraphics[width=7.0 cm]{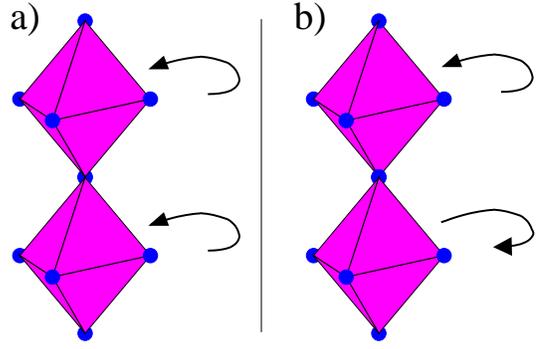}
\caption{(Color online) \label{ROTATE} The two $\theta$ modes for a bilayer.
Left: the two layers ($a$ and $b$)
rotate in phase as in Eq. (\ref{THETAB}). Right: the
two layers rotate out of phase as in Eq. (\ref{THETAA}).  The out of phase
rotation increases the energy by twisting the oxygen orbitals but this is
compensated by reducing the Coulomb interactions between octahedra.
First principles calculations[\onlinecite{TYFP}] indicate that these
modes differ only slightly in energy.}
\end{center}
\end{figure}

Thus, as shown in Fig. \ref{ROTATE}, we have two possible distorted
structures, depending on the sign of $b$:
\begin{eqnarray}
\theta_{1,a}&=&-\theta_{1,b}= \xi \theta_{2,a}= - \xi \theta_{2,b} \ ,
\hspace{0.3 in}b>0 \label{THETAA}  \\
\theta_{1,a}&=&\theta_{1,b}= \xi \theta_{2,a}=  \xi \theta_{2,b} \ ,
\hspace{0.3 in} b<0 \label{THETAB} \ ,
\end{eqnarray}
where $\xi = \pm 1$. The fact that $\xi$ can have either sign indicates
the decoupling of even and odd numbered sublattices which we mentioned
in connection with the analogous RP214 $\theta$ structures.

\begin{figure} [h!]
\begin{center}
\includegraphics[width=9.2 cm]{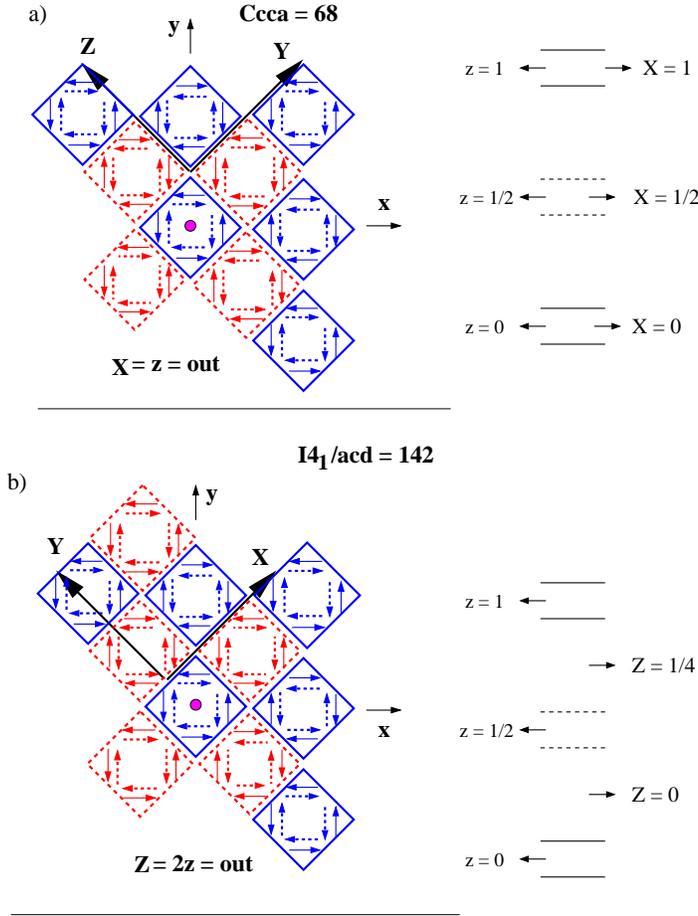}
\caption{(Color online) \label{THROT2} As Fig. \ref{THROT}. The $\theta$-modes
with $\theta_{n,a}=-\theta_{n,b}$.  The full (dashed) squares are equatorial
sections of the bilayer centered at $z=0$ ($z=1/2$).  The outer
full (inner dashed) arrows are the displacements of the equatorial oxygens
in the equatorial plane in the upper (lower) layer of the bilayer.  
In the upper (lower) panel the arrows are unchanged (reversed) under
$z \rightarrow z+1$.  The new origin is at $z=0$ in the upper panel
and at $z=1/4$ in the lower panel. At right we show the $z$-coordinates
of the full and dashed layers.}
\end{center}
\end{figure}

\begin{figure} [h!]
\begin{center}
\includegraphics[width=8.0 cm]{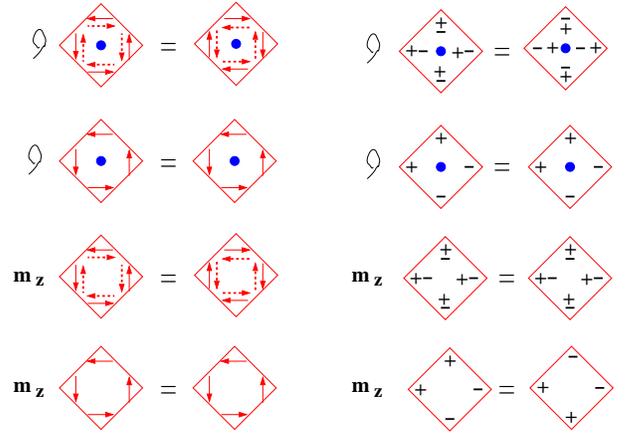}
\caption{(Color online) \label{ZSYM} The effect of operations which
take $Z$ into $-Z$.  Here the distortion of a bilayer is represented
by two sets of symbols within the square representing the equatorial
oxygens.  The outer set of symbols applies to the upper layer of the
bilayer and the inner set to the lower layer of the bilayer.
Inversion (${\cal I}$) is taken with respect to the (blue) filled
circle as the origin, which for bilayers is in the plane midway between
the upper and lower layer. $m_z$ is a mirror operation with respect to
the plane perpendicular to the tetragonal $z$-axis which passes
through the origin.}
\end{center}
\end{figure}

We now determine the space groups corresponding to these two modes.
In the mode of Eq. (\ref{THETAB}) the two layers can be coalesced
continuously into a single layer.  So this structure has the same
symmetry as the Cmca $\theta$-structure resulting from the star of
${\bf X}$ and is obtained by replacing each layer by a bilayer with
in phase rotations.  See Fig. \ref{THROT}a. The mode of Eq. (\ref{THETAA})
is shown in Fig. \ref{THROT2}a. We have to discuss the way we depict
bilayer systems in our figures.  Within each square 
(which represents the equatorial
oxygen of an octahedron) the symbols closer to the corners of the
square apply to the upper layer of the bilayer and the symbols closer
to the center apply to the lower layer of the bilayer. This convention
does not cause undue difficulty in visualizing the effect of rotations
about an axis perpendicular to the plane of the paper.  However, operations
such as reflection through the plane of the paper or inversion relative to
the center of the octahedron introduce visual complications because these
operation interchange inner and outer symbols.  Experience
indicates that for these operations one should use the results of
Fig. \ref{ZSYM}, supplemented, if need be, by a translation.
Operations such as ${\cal O}=(X, \overline Y, \overline Z)$ which involve
taking $Z$ into $-Z$ are best expressed  as ${\cal O}= {\cal I}(\overline X,
Y,Z)$ or  ${\cal O}=m_z (X, \overline Y, Z)$.  The mirrors perpendicular
to the page do not cause any confusion because for them outer
symbols are taken into outer symbols, thus avoiding visual complications.
We now return to the discussion of the mode of Eq. (\ref{THETAA})
in Fig. \ref{THROT2}a. Using, if need be, the results of Fig. \ref{ZSYM},
one sees that this structure has generators $(X\pm1/2,Y+1/2,Z)$,
$(X,Y,Z+1)$, $(\overline X, \overline Y, \overline Z)$,
$(X+1/2, \overline Y , \overline Z + 1/2)$, and $(\overline X +1/2, 
\overline Y , Z)$, which therefore is Ccca (D$_{2h}^{22}$) \#68
coming from irrep ${\bf X}_1^-$.

\subsubsection{$\phi$ structures}

Now we analyze the free energy for $\phi$ distortions at the
star of ${\bf X}$.  Here the effect of $m_z$ has the opposite
sign from the RP214 case. But since all terms are of even order,
the symmetry of the free energy is the same as for RP214.  So
\begin{eqnarray}
F &=& c_\theta a^2 \lambda \sum_{k=1}^2 \left[
\left( 2\phi_{xk}^2 + \epsilon_{xx} \right)^2
+ \left( 2\phi_{yk}^2 + \epsilon_{yy} \right)^2 \right] \nonumber \\
&& \ + c_\phi a^2 \lambda \sum_{k=1}^2 \left( 2 \phi_{xk} \phi_{yk}
+ \epsilon_{xy} \right)^2 \nonumber \\ 
&+& [(\alpha - \beta)/2] \left[ ( \phi_{x,1}-\phi_{y,2})^2
+ ( \phi_{x,2}-\phi_{y,1})^2 \right]
\nonumber \\ && + [(\alpha + \beta)/2] \left[
( \phi_{x,1}+\phi_{y,2})^2
+ ( \phi_{x,2}+\phi_{y,1})^2 \right]\   ,
\end{eqnarray}
where $\phi_{\alpha,n}$ is the value of $\phi_\alpha$ for the
top layer of the $n$th ($n=1,2$) slab.

\begin{figure} [h!]
\includegraphics[width=7.0 cm]{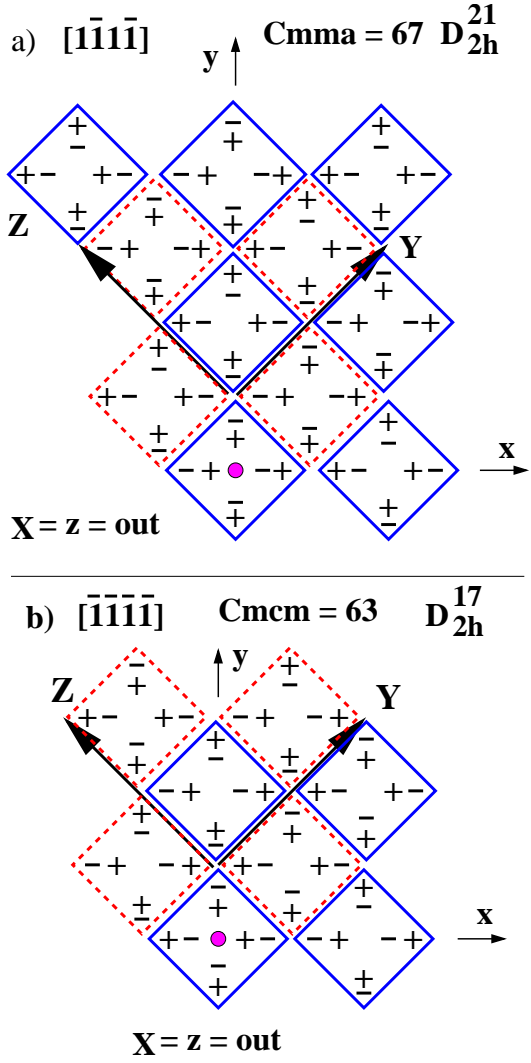}
\vspace{0.1 in}
\caption{\label{2XFIG} (Color online)
As Fig. \ref{XP} but for the star of ${\bf X}$ for RP327 (with invariance
under $z \rightarrow z+1$).  $x$, $y$, and $z$ are the tetragonal axes and
and $X$, $Y$, and $Z$ are the conventional lattice vectors after distortion.
The filled magenta circle is the tetragonal origin.  The outer $+$ or $-$
sign gives the sign of the $z$-component of displacement of the upper
layer of the bilayer and the inner $+$ or $-$ sign gives the sign of the
$z$-component of displacement of the lower layer of the bilayer.
$\Psi=$ $[1 \overline 1 1 \overline 1]$ for a) and
$[\overline 1 \overline 1 \overline 1 \overline 1]$ for b).
The new origin is in the $z=0$ plane.  The third
new axis vector is $[001]$ in terms of the original tetragonal coordinates.}
\end{figure}

The analysis parallels that for RP214 systems.  Here we characterize the
structures by the values of $\phi_{n,\alpha}$ in the {\it top} layer of
each bilayer. From the first line we conclude that $\phi_{x1}^2
= \phi_{x2}^2$ and $\phi_{y1}^2 = \phi_{y2}^2$.  There are two cases: the
first is if $\alpha-\beta$ is critical and the second is if $\alpha+\beta$
is critical.  In the first case $\phi_{x1}=-\phi_{y2}$ and $\phi_{x2}=-\phi_{y1}$.
In the second case $\phi_{x2}=\phi_{y1}$ and $\phi_{x1}=\phi_{y2}$.
Then for the first case
the possible ordering vectors are
proportional to $\Phi=[11 \overline 1 \overline 1]$ or $[1 \overline 1 1
\overline 1]$.  These are equivalent structures and we take the second
one as the representative for this case.  For the second case
the possible ordering vectors are proportional to $\Phi=
[1111]$ and $[1 \overline 1 \overline 1 1]$. These are equivalent
structures and we take $[\overline 1 \overline 1 \overline 1 \overline 1]$
as the representative for this case.  Fig.  \ref{2XFIG} shows these
representatives.

Finally, we identify the space groups of the structures of Fig.  \ref{2XFIG}.
Note that $\alpha+\beta$ critical corresponds to irrep $X_4^-$ and
$\alpha-\beta$ critical corresponds to irrep $X_3^-$ in the notation
Ref. \onlinecite{ISOTROPY}.  The generators of
$[1 \overline 1 1 \overline 1]$ are $(X\pm 1/2, Y+1/2,Z)$,$(X,Y,Z+1)$,
$(\overline X + 1/2, \overline Y, Z)$, $(X,\overline Y,\overline Z)$,
and $(\overline X , \overline Y, \overline Z)$ and those of
$[\overline 1 \overline 1 \overline 1 \overline 1]$ are
$(X \pm 1/2, Y+1/2,Z)$, $(X,Y,Z+1)$, $(\overline X , \overline Y ,\overline Z)$,
$(\overline X,\overline Y, Z+1/2)$, and $(X, \overline Y, \overline Z)$.
From these generators we identify the space groups as indicated in Fig.
\ref{2XFIG}.

\subsection{The star of {\bf N}}

Again the analysis parallels that for RP214.  Now the analog of Table 
\ref{NTAB} is Table \ref{2NTAB}.

\begin{table} [h!]
\caption{\label{2NTAB} As Table \ref{TAB1}, but for the RP327 variables of
the star of ${\bf N}\equiv (1/2,0,1/2)$ (shown in Fig. \ref{N}).
$T\equiv (1/2,1/2,1/2)$, $T_x \equiv (1,0,0)$, and $T_y \equiv (0,1,0)$.
All variables are odd under ${\bf T} \equiv (0,0,1)$.}
\vspace{0.2 in}
\begin{tabular} {|| c | c| c| c| c|c|c||}
\hline  \hline 
& ${\cal R}_4$ & $m_d$ & $m_z$ & $T$ & $T_x$ & $T_y $\\
\hline
$\phi_{x,1}$ & $\phi_{y,1}$ & $\phi_{y,1}$ & $\phi_{x,1}$ & $\phi_{x,2}$ &
$-\phi_{x,1}$ & $\phi_{x,1}$ \\ 
$\phi_{y,1}$ & $-\phi_{x,1}$ & $\phi_{x,1}$ & $\phi_{y,1}$ & $\phi_{y,2}$ &
$\phi_{x,1}$ & $- \phi_{y,1}$ \\ 
$\phi_{x,2}$ & $\phi_{y,2}$ & $\phi_{y,2}$ & $-\phi_{x,2}$ & $-\phi_{x,1}$ &
$-\phi_{x,2}$ & $\phi_{x,2}$ \\ 
$\phi_{y,2}$ & $\phi_{x,2}$ & $\phi_{x,2}$ & $-\phi_{y,2}$ & $-\phi_{y,1}$ &
$\phi_{y,2}$ & $-\phi_{y,2}$ \\ 
\hline \hline \end{tabular}
\end{table}

\begin{figure} [h!]
\begin{center}
\includegraphics[width=7.0 cm]{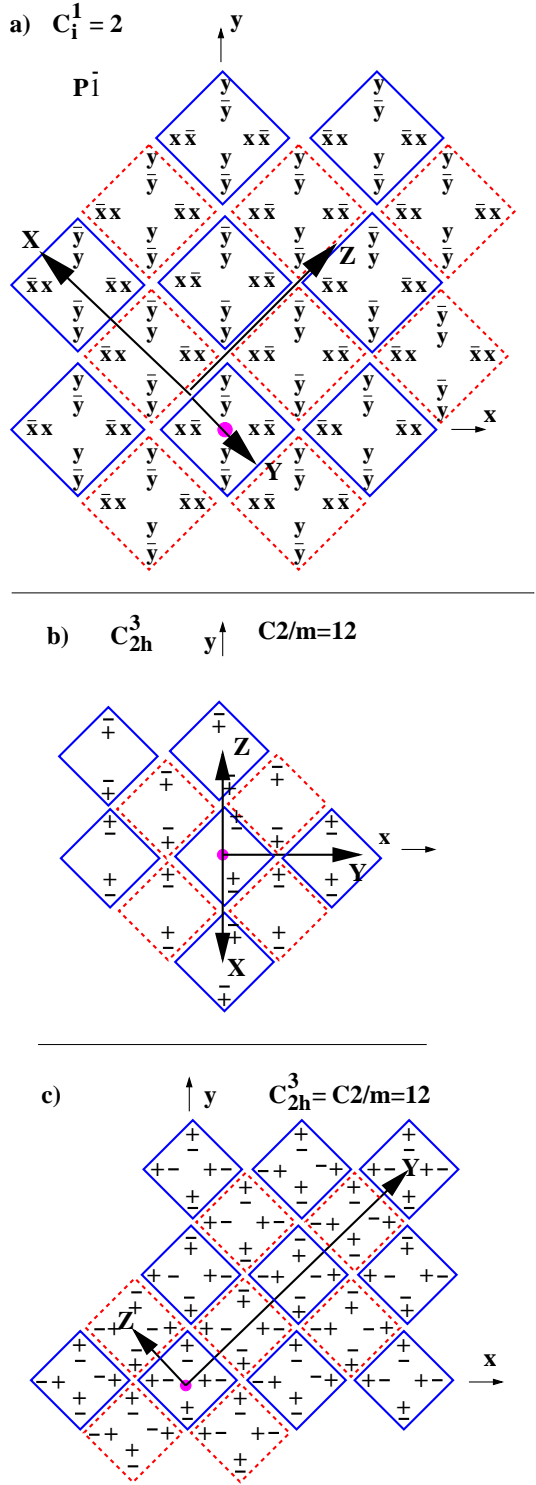}
\caption{(Color online) \label{2NFIG} As Fig. \ref{2XFIG} but for the
star of ${\bf N}$ (with sign change under $z \rightarrow z+1$) for a)
$[\overline x y \overline x \overline y]$, b) $[010\overline 1]$, and
c) $[\overline 1 1 \overline 1 \overline 1]$.  The new origins are at
$z=1/4$ for a) and at $z=1/2$ for b) and c).} 
\end{center}
\end{figure}

The free energy for the star of ${\bf N}$ is then
\begin{eqnarray}
F &=& c_\theta a^2 \lambda \sum_{k=1}^2 \left[
\left( 2\phi_{xk}^2 + \epsilon_{xx} \right)^2
+ \left( 2\phi_{yk}^2 + \epsilon_{yy} \right)^2 \right] \nonumber \\
&& + \frac{\alpha}{2} \left[ \phi_{x,1}^2  + \phi_{y,1}^2 + \phi_{x,2}^2 
+ \phi_{y,2}^2 \right] + F_{2 \epsilon} + F_4 \ . 
\label{2NEQ} \end{eqnarray}
This free energy is the same as for RP214, so the structures for RP327
will be related to those of RP214.  In Fig. \ref{2NFIG} we show the structures
in which the $\phi$'s for the upper layer are identical to those of the 
single layer in Fig. \ref{NFIG}.  Note that of all the generators of Fig.
\ref{NFIG}, only inversion $(\overline X, \overline Y, \overline Z )$ connects
the $z=n$ layers to the $z=n+1/2$ layers.  Also note that according to
Fig. \ref{ZSYM} inversion for $n=2$ systems introduces an extra minus sign
compared to the $n=1$ case.  This fact suggests that by appropriately
placing the new origins (as we have done in Fig. \ref{2NFIG}) one can pass
from RP214 to RP327, otherwise keeping the structure (and the space group)
unchanged.  We thereby determine the
generators to be the same as for the analogous structures in Fig. \ref{NFIG}:
for $[\overline x y \overline x \overline y ]$ the generators are
$(X+1,Y,Z)$, $(X,Y+1,Z)$, $(X,Y,Z+1)$, and $(\overline X, \overline Y,
\overline Z)$, which is C$_i^1$ (P$\overline 1 =$ \# 2).  For both
$[010\overline 1 ]$ and $[\overline 1 1\overline 1 \overline 1]$, the
generators are $(X,Y,Z+1)$, $(X+1/2,Y+1/2,Z)$, $(X+1/2, Y-1/2,Z)$,
$(\overline X , \overline Y , \overline Z )$, and
$(\overline X , Y , \overline Z )$ which is C$_{2h}^3$ (C2/m $=$ \#12).
Although these two structures belong to the same space group, they are
different because their unit cells are different (see Fig. \ref{NFIG}).


\subsection{The star of {\bf P}}

\subsubsection{$\theta$ structures}

The situation for $\theta$-structures at the star of ${\bf P}$ is
similar to that at the star of ${\bf X}$ except that the displacements
change sign under $z \rightarrow z+1$.  The structures with
$\theta_{n,a}=\theta_{n,b}$ have the same symmetry as that shown in
Fig. \ref{THROT}b: however, each single layer is replaced by a bilayer
in which $\theta_{n,a}=\theta_{n,b}$.  So the space group is again
I$4_1$/acd=\#142 (D$_{4h}^{20}$). The structure with 
$\theta_{n,a}=-\theta_{n,b}$ and which changes sign under $z \rightarrow z+1$
is illustrated in Fig. \ref{THROT2}b. This structure is generated by
$(X-1/2,Y+1/2,Z+1/2)$, $(X+1/2,Y-1/2,Z+1/2)$, $(X+1/2,Y+1/2,Z-1/2)$,
($\overline X , \overline Y, \overline Z)$,
$(X,\overline Y, \overline Z + 1/2)$, and $(\overline Y +1/4, X+3/4,Z+1/4)$
and is therefore I4$_1$/acd =  \#142 (D$_{4h}^{20})$ coming from irrep
P$_2$.  In identifying the generators we used Fig. \ref{ZSYM}.

\subsubsection{$\phi$ structures}

\begin{figure} [h!]
\begin{center}
\includegraphics[width=8.6 cm]{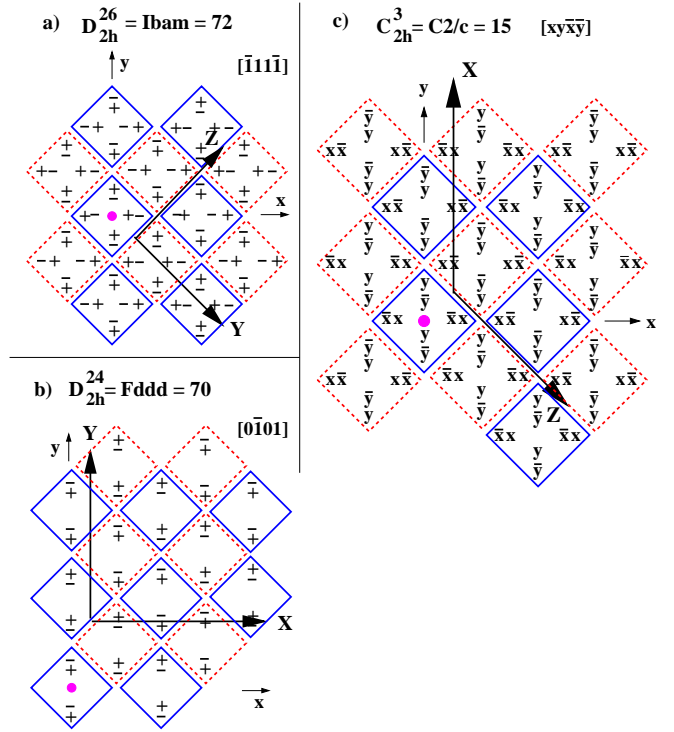}
\caption{(Color online) \label{2PFIG} As Fig. \ref{2XFIG} but for the
star of ${\bf P}$ (with sign change under $z \rightarrow z+1$) for
a) $[1\overline 1 1 \overline 1]$, b) $[0 \overline 1 0 1]$,
and c) $[xy \overline x \overline y ]$.
The new origins are at $z=3/4$ in a) and in b) and at $z=1/4$
in c).  The new out-of-plane axes are $(0,0,2)_t$.}
\end{center}
\end{figure}

To identify the space groups of the structures shown in Fig. \ref{2PFIG} recall
the discussion just below Eq. (\ref{2NEQ}).  By shifting the new origin, we
identify the generators as:
$(X, Y, \overline Z)$,[\onlinecite{IBAM}]
$(X, \overline Y, \overline Z +1/2)$, $(\overline X, \overline Y, \overline Z )$,
$(X-1/2,Y+1/2,Z+1/2)$, $(X+1/2,Y-1/2,Z+1/2)$, and $(X+1/2,Y+1/2,Z-1/2)$ for a),
$(\overline X + 1/4, \overline Y + 1/4,Z)$, $(X, \overline Y +1/4, \overline Z +1/4)$,
$(\overline X, \overline Y, \overline Z)$,
$(X-1/2,Y+1/2,Z+1/2)$, $(X+1/2,Y-1/2,Z+1/2)$, and $(X+1/2,Y+1/2,Z-1/2)$ for b), and
$(\overline X , Y, \overline Z)$, $(\overline X ,\overline Y, \overline Z )$,
$(X+1,Y,Z)$, $(X,Y+1,Z)$, and $(X,Y,Z+1)$ for c).
These lead to the space groups listed in the figure. 
Because of the Lifshitz 
instability the wave vector is an incommensurate one close to ${\bf P}$,
as is discussed in Sec. III.2.

\section{RP Systems with $n>2$}

Now we are in a position to analyze the situation of RP systems with
$n>2$. For the $\theta$-structures the important issue is whether the
eigenvalue $\lambda(M_z)$ which gives the symmetry of the stacking sequence
is $+1$ or $-1$.  For the $\phi$-structures the important issue is, as stated 
in Ref. \onlinecite{PHASE}, whether $n$, the number of layers per substructure,
is even or odd.  

\subsection{$\theta$-structures}


\begin{figure} [h!]
\begin{center}
\includegraphics[width=6.5 cm]{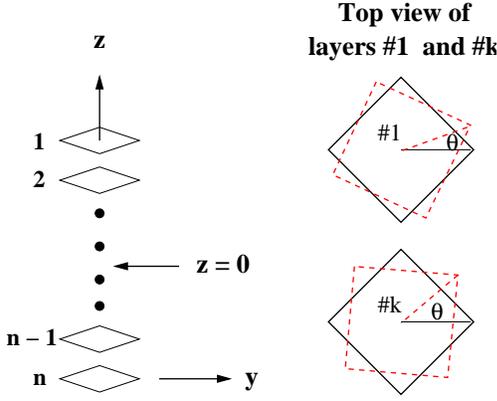}
\caption{(Color online) \label{COORDS} The $\theta$-structures for one of the
$n$-layer slabs.  Left: The cross section of the $n$-layer slab showing the
$(x,y,z)$ coordinate system used to discuss the symmetry of the
$\theta$-structures.  Right: Top view showing that the structure (red dashed
cross section) is characterized by giving the value of the rotation $\theta$
for each of the $n$ layers.}
\end{center}
\end{figure}

We first consider $\theta$-structures.  The $\theta$-structures associated
with the star of either ${\bf X}$ or of ${\bf P}$ are governed by the free
energy
\begin{eqnarray}
F &=& \sum_{m=1}^2 \sum_{k =1}^n \left[
\left( 2 \theta_{mk}^2 - \epsilon_{xx} \right)^2 +
\left( 2 \theta_{mk}^2 - \epsilon_{yy} \right)^2 \right]
\nonumber \\ && \ 
+ \sum_{m=1}^2 \sum_{k,l=1}^n A_{k,l} \theta_{m,k} \theta_{m,l}
+ V \ ,
\label{AAEQ} \end{eqnarray}
where $V$ contains interaction terms (of order $\lambda^0$) 
between the two different slabs. Note that quadratic terms
like $\theta_{1,\alpha} \theta_{2,\beta}$ are excluded because they
are not invariant under ${\cal R}_4$, as one sees from Table \ref{TAB327}
or Fig. \ref{THETA}.
The ordering vector of the first $n$-layer slab in the unit cell is
\begin{eqnarray}
\Phi_1 &=& \left[ \theta_{1,1} , \theta_{1,2}, \theta_{1,3} \dots \theta_{1,n}
\right] \ .
\label{EQ33} \end{eqnarray}
Note that $\Phi$ is proportional to the eigenvector of the matrix
${\bf A}$ which has the minimal eigenvalue (so that it is the one
which first becomes critical as the temperature is lowered).
The matrix ${\bf A}$ is invariant under the mirror operation
$M_z$ which interchanges layers $k$ and $n+1-k$,
so that $\theta_k \leftrightarrow \theta_{n+1-k}$.  Therefore
we know only that the eigenvector is either even or odd ({\it i. e.} the
eigenvalue of $M_z$, $\lambda (M_z)$, is either $+1$ or $-1$,
respectively), depending on the details of the interactions in the system.
For instance, for $n=1$ $\Phi_1=[1]$ and is even under $M_z$.  For $n=2$ the
critical eigenvector is either $[1 1]$ (which is even under $M_z$) or
$[1 \overline 1]$ (which is odd under $M_z$).  For $n=3$ the critical
eigenvector is either $[1 0 \overline 1]$ (which is odd under $M_z$) or
$[\alpha \beta \alpha ]$, where $\alpha$ and $\beta$ depend on the
interactions within the 3-layer subsystem and this eigenvector is even under
$M_z$.  On the basis of symmetry we can definitely {\it not} posit any
specific form for the eigenvector (for $n>2$), as is done in Table XIII of
Ref.  \onlinecite{PHASE}.  In all these examples the ordering vector for the
second $n$-layer subsystem obeys $\Phi_2 = \pm \Phi_1$, where
the indeterminancy in sign reflects the by now familiar frustration
of $\theta$ structures.

\begin{figure} [h!]
\begin{center}
\includegraphics[width=6.0 cm]{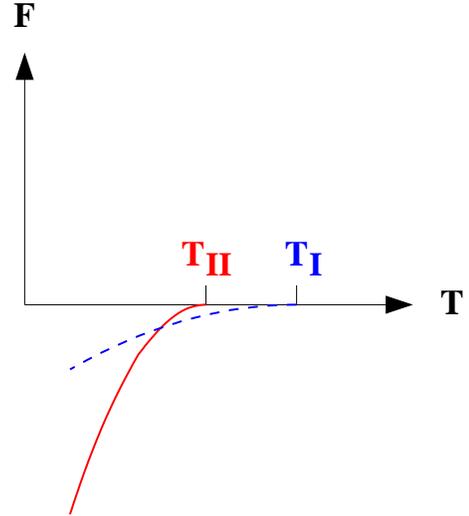}
\caption{(Color online) \label{COMPETE} The free energies of phase I (dashed
line) and phase II (full line), as given by Eq. (\ref{COMPEQ}). These curves are
drawn for $\alpha \lambda / \beta =9$.}
\end{center}
\end{figure}

For $n>2$ the state we call I, given by the critical eigenvector of ${\bf A}$,
will develop at the structural phase transition at a critical temperature
we denote $T_I$.  For this state $\theta_k^2$
will not be independent of $k$ unless there is an unusual accidental
degeneracy and therefore its free energy will be of the form
\begin{eqnarray}
F_I &=& \frac{1}{2} (T-T_I) Q_I^2 + \alpha \lambda Q_I^4 \ .
\end{eqnarray}
The quartic term must be of order $\lambda$ because $\theta_k^2$ is
{\it not} independent of $k$.
This state competes with state II which satisfies (for all $k$ and $l$)
\begin{eqnarray}
2 \theta_{kl}^2 &=&  - \epsilon_{xx} = - \epsilon_{yy} \ .
\end{eqnarray}
State I becomes critical at a temperature $T_I$ which is higher than that,
$T_{II}$, at which state II becomes critical. The free energy of
state II can be written as
\begin{eqnarray}
F_{II} &=& \frac{1}{2} (T-T_{II}) Q_{II}^2 + \beta Q_{II}^4 \ . 
\end{eqnarray}
Minimization with respect to the order parameters yields
\begin{eqnarray}
F_I &=& - \frac{1}{2} \frac{(T-T_I)^2}{\alpha \lambda} \hspace{0.2 in}
T < T_I \ , \hspace{0.4 in} F_I=0 \ , \hspace{0.2 in} T > T_I \nonumber \\
F_{II} &=& - \frac{1}{2} \frac{(T-T_{II})^2}{\beta} \hspace{0.2 in}
T < T_{II} \ , \hspace{0.4 in} F_{II}=0 \ , \hspace{0.2 in} T > T_{II} \ .
\label{COMPEQ} \end{eqnarray}
and these are plotted versus $T$ in Fig. \ref{COMPETE}.  One sees that
for $n>2$ we have a more complicated phase diagram than for $n=1$ or 2.
When $\lambda$ is large, for a small range of temperature phase I is stable,
but at lower $T$ we arrive at phase II.  For $n>3$ it is possible to have
more than one phase transition before ultimately reaching phase II.

\begin{figure} [h!]
\begin{center}
\includegraphics[width=6.0 cm]{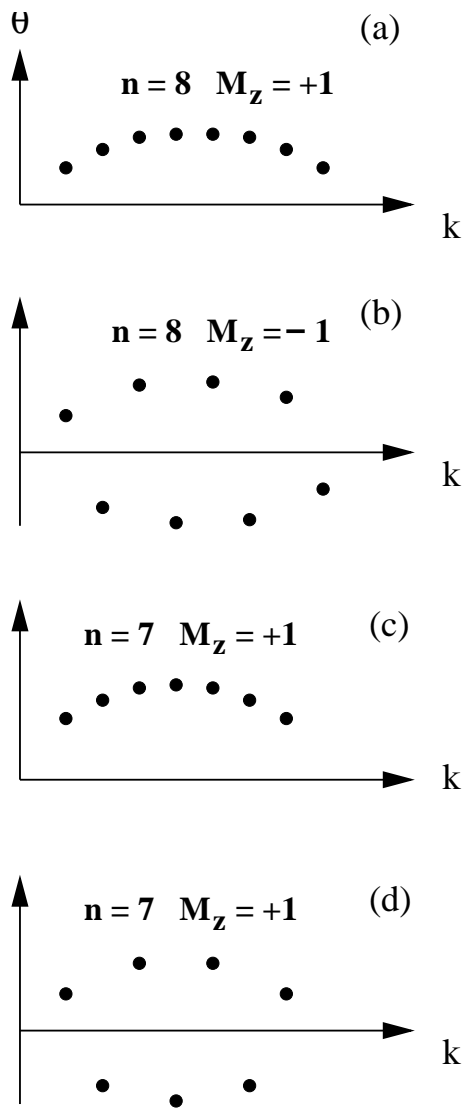}
\caption{\label{GRAPH} Typical variation of $\theta$ as a function of 
the layer index, $k$, for dominant nearest neighbor interlayer interaction,
$A_{k,k+1} \equiv w$.  For case (a), $n$ is even and $w <0$, so that
$\lambda(M_z)=+1$. For case (b), $n$ is even and $w > 0$, so that
$\lambda(M_z)=-1$.  For cases (c) and (d) $n$ is odd. For case (c)
$w<0$ and for case (d) $w>0$.  In both cases $\lambda(M_z)=+1$.  To obtain
$\lambda(M_z)=-1$ for $n$ odd requires further neighbor interlayer
interactions.}
\end{center}
\end{figure}
 
The possible $\theta$-structures of phase I depend on the wave vector
(${\bf X}$ or ${\bf P}$) and whether the interactions within the system
select $\lambda(M_z)=+1$ or $-1$.  First consider the star of ${\bf X}$,
for which $\theta_{m+2,k}=\theta_{m,k}$.  If
$\lambda(M_z)=+1$ (as in Fig. \ref{THROT}a), then, irrespective of the value
of the number of layers per subsystem, we will have a structure similar
to that of Fig. \ref{THROT}a (in which each layer of Fig. \ref{THROT} is
replaced by an $n$-layer slab) with space group Cmca=\#64. If interactions
select, $\lambda(M_z)=-1$ (this is only possible for $n>1$), then we have a
structure of space group Ccca=\#68, similar to that shown in Fig.
\ref{THROT2}a. Next consider the star of ${\bf P}$, for which $\theta_{m+2,k}
=-\theta_{m,k}$.
If $\lambda(M_z)=+1$ (as in Fig. \ref{THROT}b), then, irrespective of the value
of the number of layers per subsystem, we will have a structure similar
to that of Fig. \ref{THROT}b with space group I4$_1$/acd=\#142. If interactions
select $\lambda(M_z)=-1$ (this is only possible for $n>1$), then we have a
structure of space group I4$_1$/acd=\#142, similar to that shown in Fig.
\ref{THROT2}b. 

The above remarks relied only on symmetry.  However, now we consider
the likely form of the interaction matrix $A_{k,l}$ in Eq. (\ref{AAEQ})
which determines $\theta$ as a function of the layer index $k$.
If the dominant intraslab interactions are those between adjacent
layers, then, as illustrated in Fig. \ref{GRAPH}, we obtain configurations
analogous to ferromagnetic (panel a) or antiferromagnetic (panel b)
spin structures.  Thus, with nearest neighbor interlayer interactions,
if $n$, the number of layers per slab is even, we can have
either sign of $\lambda(M_z)$.  If $n$ is odd, then this special ansatz of
nearest neighbor interlayer interactions can only give $\lambda(M_z)=+1$.
If, experimentally, the case $\lambda(M_z)=-1$ is observed for odd $n$,
one could conclude the existence of significant longer ranger 
interlayer interactions.  (Such a situation is obviously possible
in the presence of Coulomb interactions.)

Our results are summarized in Table \ref{TOTAL}. Note that for the
$\theta$ structures the controlling variable is not the number of
layers $n$, but $\lambda (M_z)$.  It is interesting to note that for
${\bf N}$ and ${\bf P}$, the $\phi$-structures for even and odd $\lambda (M_z)$ 
are very similar.  Apart from the fact that their substructures are different,
they only differ in the location of the center of inversion symmetry.

\begin{table}
\caption{\label{TOTAL} Summary of results for commensurate structures
for RP $n$ layer systems. $\vec Q$ denotes the wave vector, Var labels
the angular variable, and $\lambda$
is the eigenvalue of the mirror operation within the $n$-layer substructure,
as discussed below Eq. (\ref{EQ33}).  In the last column we indicate whether
the transition is allowed to be continuous (Y) or not (N) or whether there
is a Lifshitz instability (L). See Ref. \onlinecite{ISOTROPY}. The results
in this table can be compared to the results for the $X$ point given in
Table XIII of Ref. \onlinecite{PHASE}.  Our analysis only allows
structures \#2, 5, 9, and 12 of that reference.}
\vspace{0.2 in}
\begin{tabular} {|| c c c | c c c| c|| }
\hline \hline
$\vec Q$ & Var & \ $\lambda$ \ & Space group(s) & $n$ & See Fig. & Y,N,L \\
\hline
${\bf X}$ & $\theta$ & $-1$ & \#68 (D$_{2h}^{22}$) Ccca & $n>1^*$ & 14a & Y\\
${\bf X}$ & $\theta$ & $+1$ & \#64 (D$_{2h}^{18}$) Cmca & $n \geq 1$ & 7a& Y \\
\hline \hline
${\bf P}$ & $\theta$ & $+1$ & \#142 (D$_{4h}^{20}$) I4$_1$/acd & $n \geq 1$ &
7b & Y\\
${\bf P}$ & $\theta$ & $-1$ & \#142 (D$_{4h}^{20}$) I4$_1$/acd & $n > 1^*$
& 14b& Y\\
\hline \hline
${\bf X}$ & $\phi$ & $-1$ & \#64 (D$_{2h}^{18}$) Cmca &  odd & 8a& Y \\ 
&&& \#66 (D$_{2h}^{20}$) Cccm &  odd & 8b& Y \\
${\bf X}$ & $\phi$ & $+1$ & \#63 (D$_{2h}^{17}$) Cmcm &  even & 16c& Y \\
&&& \#67 (D$_{2h}^{21}$) Cmma &   even & 16a& Y \\
\hline \hline
${\bf N}$ & $\phi$ & $-1$ & \#12 (C$_{2h}^3$) C2/m &  odd & 10b&N \\
&&& \#12 (C$_{2h}^3$) C2/m &   odd & 10c&N \\
& & & \#2 (C$_i^1$) P$\overline 1$ &  odd & 10a&N \\
${\bf N}$ & $\phi$ & $+1$ & \#12 (C$_{2h}^3$) C2/m &  even & 17b&N \\
&&& \#12 (C$_{2h}^3$) C2/m  &  even & 17c&N \\
&&& \#2 (C$_i^1$) P$\overline 1$ &  even & 17a&N \\
\hline \hline 
P & $\phi$ & $-1$ & \#72 (D$_{2h}^{26}$) Ibam & odd & 11a &L \\ 
&&& \#70 (D$_{2h}^{24}$) Fddd & odd & 11b &L \\ 
&&& \#15 (C$_{2h}^3$) C2/c & odd & 11c &L \\ 
P & $\phi$ & $+1$ & \#72 (D$_{2h}^{26}$) Ibam & even &  18a &L \\ 
&&& \#70 (D$_{2h}^{24}$) Fddd & even & 18b &L \\ 
&&& \#15 (C$_{2h}^3$) C2/c & even &18c  &L \\ 
\hline \hline 
\end{tabular}

\noindent $^*$
If the dominant interlayer interactions are those between adjacent layers,
then, as discussed in the text, $n$ must be even for this case to occur.
\end{table}

\section{DISCUSSION AND CONCLUSION}

We did not deal with the positions of the ions at the center of
the octahedra or those between the layers of octahedra.  Here we
are only concerned with such ionic positions as they are
modified by the orientational structural transition.  Each such
ion sits in a stable potential well.  The question is whether or not
for systems without any accidental degeneracy there is a bifurcation
so that additional space groups could be allowed when the positions
of these ``inessential" ions are taken into account.  The stable
potential well can be distorted and the placement of its minimum
will be modified by the octahedral reorientation.  But a single
minimum of a stable potential well can not be continuously deformed
into a double well without assuming an accidental vanishing of the
fourth order term in the local potential.  Similar arguments show
that the perturbative effect of the center of mass coordinates
of the nearly rigid octahedra do not produce anomalous effects. 
Of course, parameters of Wykoff orbits which are not fixed by symmetry
will be perturbatively modified at the structural phase transition.
Similarly, the elastic strain tensor will be perturbatively modified
consistent with the symmetry of the resulting phase at the transition.

Experimentally, it is striking that the structures observed as distortions
from the tetragonal phase are in our much shorter list.  For instance,
in the data cited on p 313 and ff of Ref. \onlinecite{PHASE} five systems
with $\phi$ tilts are shown which go into either Cmca (64) or 
P4$_2$/ncm (138), except for Rb$_2$CdCl$_4$ whose structure is
uncertain: either Cmca or Fccm (which is on neither our list
nor that of Ref. \onlinecite{HANDS} because it involves two irreps).
Systems (other than Rb$_2$CdCl$_4$ subsequently discussed in Ref.
\onlinecite{PHASE}) in Table III of Ref. \onlinecite{HANDS} likewise
go into either Cmca or P4$_2$/ncm.

To summarize: we have analyzed the possible structural transitions of
the so-called Ruddlesden-Popper perovskite structure (such as 
K$_2$MnF$_4$ or Ca$_3$Mn$_2$O$_7$, etc.)
using a variant of Landau theory in which the constraint of rigid
oxygen octahedra is implemented and our results are compared to the
well-known results of Refs. \onlinecite{HANDS} and \onlinecite{PHASE}.
A check on the accuracy of our treatment of symmetry is that our
list of allowed structures (for K$_2$MgF$_4$) which can be reached via
a single structural phase transition is a {\it subset} of the list of
Ref. \onlinecite{HANDS}.  We find that the rigid
octahedral constraint eliminates all the structures in Table I of Ref.
\onlinecite{HANDS} for which the octahedral tilting transitions are
discontinuous. It is also appealing that structures which are
allowed by symmetry but which involve undistorted sublattices are
eliminated by the octahedral constraint. The results for the K$_2$MgF$_4$
structure are summarized
in Tables \ref{NO} and \ref{YES}, where one sees that our list of
possible structures is a much reduced subset of Ref. \onlinecite{HANDS}.
For these systems, our analysis allows 
(see Tables \ref{NO} and \ref{YES}) only 13 of the 41 structures
listed in Ref. \onlinecite{HANDS} and of these only nine are commensurate
structures.  A summary of our results for commensurate structures for
A$_{n+1}$B$_n$C$_{3n+1}$ is given in Table \ref{TOTAL}.

\noindent
ACKNOWLEDGEMENTS: I am grateful to T. Yildirim for performing the first
principles calculations.  I thank
J. M. Perez-Mato, C. J. Fennie, B. Campbell, and H. T. Stokes
for helpful discussions.

\begin{appendix}
\section{Minimization to order $\lambda^{-1}$}

Here we show that corrections of order $1/\lambda$ to the minimization 
of the free energy do not qualitatively affect our conclusions.  We consider
the free energy
\begin{eqnarray}
F &=& c_\theta a^2 \lambda 
\left[ \left( \frac{1}{2} \theta_1^2 + \epsilon_{xx} \right)^2
+ \left( \frac{1}{2} \theta_2^2 + \epsilon_{xx} \right)^2 \right. \nonumber \\
&& \ + \left. \left( \frac{1}{2} \theta_1^2 + \epsilon_{yy} \right)^2
+ \left( \frac{1}{2} \theta_2^2 + \epsilon_{yy} \right)^2 \right]
\nonumber \\ && \
- \frac{1}{2} | \gamma| [\theta_1^2 + \theta_2^2] 
+ \frac{1}{2} \sum_{j,k} c_{j,k} \epsilon_j \epsilon_k
\nonumber \\ && \
+ \frac{1}{4} u [\theta_1^4 + \theta_2^4] + \frac{1}{2} v \theta_1^2 \theta_2^2
- d \epsilon_{xy} \theta_1 \theta_2 \nonumber \\ && \
+ \left[ b \left( \epsilon_{xx}+\epsilon_{yy}\right) + c c_{zz} \right]
\left[ \theta_1^2 + \theta_2^2 \right] 
\end{eqnarray}
where all the coefficients except $\lambda$ are of order unity and the
summation is in Voigt notation. We assume that nonlinear elastic terms
(of higher than quadratic order) in $\epsilon$ can be neglected.  We now set
\begin{eqnarray}
\epsilon_{xx} &=& - \frac{1}{4} [ \theta_1^2 + \theta_2^2] + \xi_x \nonumber \\
\epsilon_{yy} &=& - \frac{1}{4} [ \theta_1^2 + \theta_2^2] + \xi_y \ .
\end{eqnarray}
The terms involving $\lambda$ become
\begin{eqnarray}
F(\lambda) &=& \frac{1}{4} c_\theta a^2 \lambda 
\left( \theta_1^2 - \theta_2^2\right)^2 
+ 2 c_\theta a^2 \lambda [\xi_x^2+\xi_y^2] \ .
\end{eqnarray}
Therefore the quartic terms in $\theta_k$ are of the form of Eq. (\ref{EQCC})
where $v$ is surely negative for large $\lambda$. Accordingly, to all
orders in $1/\lambda$ we may set
\begin{eqnarray}
\theta_1^2 = \theta_2^2 \equiv  \theta ( \lambda ) \ .
\end{eqnarray}
Also we see that the coefficients of terms linear in $\epsilon_{xx}$ and
$\epsilon_{yy}$ are equal.  So we may set $\xi_x=\xi_y\equiv \xi$.
Then $\xi$ is determined to leading order in $1/\lambda$ by
\begin{eqnarray}
F &=& 4 c_\theta a^2 \lambda \xi^2 - |\gamma|\theta^2 + U \theta^4
+ \frac{1}{2} c_{33} \epsilon_{zz}^2
\nonumber \\ && + [(4b - x_{11}) \xi +2 c \epsilon_{zz} ] \theta^2
+ \frac{1}{2} c_{44} \epsilon_{xy}^2 \nonumber \\ && \
-d \epsilon_{xy} \theta_1 \theta_2 \ ,
\label{A5} \end{eqnarray}
where
\begin{eqnarray}
U &=& \frac{1}{2} (u+v) -2b + \frac{1}{4} c_{11} \ . 
\end{eqnarray}
Then, by minimizing $F$ of Eq. (\ref{A5}) with respect to $\xi$, we get $\xi \sim \lambda^{-1}$ or
\begin{eqnarray}
\epsilon_{xx} &=& \epsilon_{yy} = - \frac{1}{2} \theta^2 + e \lambda^{-1} \ ,
\end{eqnarray}
where $e$ is of order $\lambda^0$.  This tells us that the octahedral mismatch
(or distortion) is not zero, but is of order $\lambda^{-1}$.
The corrections of order $1/\lambda$ to the elastic constants,
do not, of course affect the symmetry of the structure.
\end{appendix}


\begin{thebibliography} {99}
\bibitem{FN1}
Lengths are in units of lattice constants $a_i$ and wave
vectors in units of $2 \pi /a_i$.
\bibitem{AXES}
Since we focus on the displacements of ions, we denote the
displacements as $\theta$ and $\phi_\alpha$ rather than
$a\theta/2$ and $a\phi_\alpha/2$, as we would if $\theta$
and $\phi_\alpha$ were the angle variables. In addition, note
that $\phi_x$ ($\phi_y$) originates from a rotation of magnitude
$\phi_x/2$ ($\phi_y/2$) about the $y$ ($x$) axis.
\bibitem{FN4}
Our notation is such that $(X',Y',Z')$ denotes the operation which
takes $(X,Y,Z)$ into $(X',Y',Z')$. Thus ${\cal R}_4 = (-Y,X,Z)$.
\bibitem{FN34}
Equivalent structures are those related by rotations and translations.
Structures related only by improper rotations (this includes mirrors)
may be enantiomorphs and may be from distinct space groups.  So here
and below we prefer to not invoke improper rotations to test for equivalence.
\bibitem{FN35}
Note: here and below (except in panel b of Fig. \ref{2PFIG}
where we make a more convenient choice of axes)
we choose the representative to reproduce the setting
chosen in Refs. \onlinecite{HANDS} and \onlinecite{ISOTROPY}.
\bibitem{IBAM}
It is more convenient to replace $(\overline x, \overline y, z)$ in the
set of generators of Ibam given in Ref. \onlinecite{ISOTROPY} 
by $(x,y,\overline z)$.
\end{thebibliography}
\end{document}